\documentclass{aa}
\usepackage{epsfig}
\usepackage{natbib}
\bibpunct{(}{)}{;}{a}{}{,}

\newbox\grsign \setbox\grsign=\hbox{$>$} \newdimen\grdimen \grdimen=\ht\grsign
\newbox\simlessbox \newbox\simgreatbox \newbox\simpropbox \newbox\wtildebox 
\setbox\simgreatbox=\hbox{\raise.5ex\hbox{$>$}\llap
     {\lower.5ex\hbox{$\sim$}}}\ht1=\grdimen\dp1=0pt
\setbox\simlessbox=\hbox{\raise.5ex\hbox{$<$}\llap
     {\lower.5ex\hbox{$\sim$}}}\ht2=\grdimen\dp2=0pt

\newcommand{\be}{\mbox{\begin{equation}}}
\newcommand{\ee}{\mbox{\end{equation}}}

\newcommand{\tdis}{\mbox{$t_{\rm dis}^{\rm total}$}}

\newcommand{\Cref}{\mbox{$m_{\rm ref}$}}

\newcommand{\msun}{\mbox{M$_\odot$}}

\renewcommand{\d}{{\rm d}} 

\begin{document}

\title{The evolution of the stellar mass function in star clusters\thanks{The models presented in this paper will be publicly available in electronic form at the CDS via anonymous ftp to \texttt{http://cdsweb.u-strasbg.fr/} (130.79.128.5) and also at \texttt{http://www.astro.uu.nl/\~{}kruijs}.}
}

\author{J.~M.~Diederik~Kruijssen \inst{1,2}}

\institute { 
                 {\inst{1}Astronomical Institute, Utrecht University, 
                 PO Box 80000, NL-3508TA Utrecht, The Netherlands\\
                 (e-mail: {\tt kruijssen@astro.uu.nl})}\\
                 {\inst{2}Leiden Observatory, Leiden University,
                 PO Box 9513, NL-2300RA Leiden, The Netherlands}
            }

\date{Received 20 September 2009 / Accepted 23 October 2009}


\abstract{The dynamical ejection of stars from star clusters affects the shape of the stellar mass function (MF) in these clusters, because the escape probability of a star depends on its mass. This is found in $N$-body simulations and has been approximated in analytical cluster models by fitting the evolution of the MF. Both approaches are naturally restricted to the set of boundary conditions for which the simulations were performed.}
{The objective of this paper is to provide and to apply a simple physical model for the evolution of the MF in star clusters for a large range of the parameter space. It should also offer a new perspective on the results from $N$-body simulations.}
{A simple, physically self-contained model for the evolution of the stellar MF in star clusters is derived from the basic principles of two-body encounters and energy considerations. It is independent of the adopted mass loss rate or initial mass function (IMF), and contains stellar evolution, stellar remnant retention, dynamical dissolution in a tidal field, and mass segregation.}
{The MF evolution in star clusters depends on the disruption time, remnant retention fraction, initial-final stellar mass relation, and IMF. Low-mass stars are preferentially ejected after $t\sim 400$~Myr. Before that time, masses around 15---20\% of the maximum stellar mass are lost due to their rapid two-body relaxation with the massive stars that still exist at young ages. The degree of low-mass star depletion grows for increasing disruption times, but can be quenched when a large fraction of massive remnants is retained. The highly depleted MFs of certain Galactic globular clusters are explained by the enhanced low-mass star depletion that occurs for low remnant retention fractions. Unless the retention fraction is exceptionally large, dynamical evolution always decreases the mass-to-light ratio. The retention of black holes reduces the fraction of the cluster mass in remnants because white dwarfs and neutron stars have masses that are efficiently ejected by black holes.}
{The modeled evolution of the MF is consistent with $N$-body simulations when adopting identical boundary conditions. However, it is found that the results from $N$-body simulations only hold for their specific boundary conditions and should not be generalised to all clusters. It is concluded that the model provides an efficient method do understand the evolution of the stellar MF in star clusters under widely varying conditions.}


\authorrunning{J.~M.~D.~Kruijssen}
\titlerunning{The evolution of the stellar mass function in star clusters}

\maketitle


\section{Introduction} \label{sec:intro}
The evaporation of star clusters is known to change the shape of the underlying stellar mass function\footnote{Hereafter, `mass function' is referred to as `MF'.} \citep{henon69,chernoff90,vesperini97b,takahashi00,portegieszwart01,baumgardt03}. This phenomenon has been used to explain the observed MFs in globular clusters \citep{richer91,demarchi07,demarchi07b}, which are flatter than typical initial mass functions \citep[IMFs, e.g.][]{salpeter55,kroupa01}. In addition, the effect of a changing MF on cluster photometry has been investigated \citep{lamers06,kruijssen08,anders09}. This has been shown to explain the low mass-to-light ratios of globular clusters \citep{kruijssen08b,kruijssen09} and to have a pronounced effect on the inferred globular cluster mass function \citep{kruijssen09b}.

The existing parameterised cluster models that incorporate a description of low-mass star depletion are restricted by the physically self-contained models on which they are based. Some studies \citep{lamers06,kruijssen08} assume an increasing lower stellar mass limit to account for the evolving MF, others \citep{anders09} fit a changing MF slope to $N$-body simulations. In both cases, the models are accurate for a certain range of boundary conditions, but they do not include a physical model and are therefore lacking flexibility. While $N$-body simulations do include the appropriate physics, they are very time-consuming. As a result, only a limited number of clusters can be simulated and the applicability of the simulations is thus restricted to the specific set of boundary conditions for which they have been run.

It would be desirable to obtain a simple physical model for the evolution of the MF, which would have a short runtime and could be used independently of $N$-body simulations. Forty years ago, a pioneering first approach to such a model was made by \citet{henon69}, who considered the stellar mass-dependent escape rate of stars from star clusters. However, the applicability of his model was limited due to a number of assumptions that influenced the results. First of all, \citet{henon69} assumed that the clusters exist in isolation and neglected the tidal field. As a consequence, the ejection of a star could only occur by a single, close encounter and the repeated effect of two-body relaxation was not included. Secondly, the {distribution of stars was independent of stellar mass, i.e.} mass segregation was not included. Both mass segregation and the influence of a tidal field are observed in real clusters, and can be expected to affect the evolution of the MF.

The aim of this paper is to derive a physical description of the evolution of the stellar MF in star clusters, alleviating the assumptions that were made by \citet{henon69}. This should explain the results found in $N$-body simulations and observations, while providing the required flexibility to explore the properties of star clusters with simple, physically self-contained models. The outline of this paper is as follows. In Sect.~\ref{sec:model}, total mass evolution of star clusters is discussed. A recipe for the evolution of the MF is derived in Sect.~\ref{sec:mf}, covering stellar evolution, the retain of stellar remnants, dynamical dissolution and mass segregation. The model is compared to $N$-body simulations in Sect.~\ref{sec:comp}. In Sect.~\ref{sec:results}, the model is applied to assess the evolution of the MF for different disruption times and remnant retention fractions. The consequences for other cluster properties are also considered. This paper is concluded with a discussion of the results and their implications.

\section{The mass evolution of star clusters} \label{sec:model}
The mass of star clusters decreases due to stellar evolution and dynamical dissolution. This is expressed mathematically as
\begin{equation}
\label{eq:dmdttot}
  \frac{\d M}{\d t} = \left(\frac{\d M}{\d t}\right)_{\rm ev} + \left(\frac{\d M}{\d t}\right)_{\rm dis} ,
\end{equation}
with $M$ the cluster mass, and the subscripts `ev' and `dis' denoting stellar evolution and dynamical dissolution. The contribution of stellar evolution to the mass loss is derived from the decrease of the maximum stellar mass with time and depends on the adopted stellar evolution model. 

The dynamical evaporation of star clusters is increasingly well understood. Over the past years it has become clear that clusters lose mass on a disruption timescale $t_{\rm dis}$ that is proportional to a combination of the half-mass relaxation time $t_{\rm rh}$ and the crossing time $t_{\rm cr}$ as $t_{\rm dis}\propto t_{\rm rh}^{x}t_{\rm cr}^{1-x}$ \citep[e.g.][]{baumgardt01,baumgardt03,gieles08}. It is found that $x=0.75$---0.80, depending on the concentration ($c= \log{(r_{\rm t}/r_{\rm c})}$) or King parameter ($W_0$) of the cluster \citep{baumgardt03}. This proportionality leads to a disruption timescale that scales with the present day mass as \citep{lamers05}:
\begin{equation}
\label{eq:tdis}
  t_{\rm dis}=t_0 M^\gamma ,
\end{equation}
with $M$ the cluster mass, $t_0$ the dissolution timescale parameter which sets the rapidity of dissolution and depends on the cluster environment, and $\gamma$ a constant related to $x$. \citet{lamers09} find $\gamma=0.62$ for $W_0=5$ and $\gamma=0.70$ for $W_0=7$. This timescale implies a mass loss rate due to dissolution that can be described with the simple relation
\begin{equation}
\label{eq:dmdt}
  \left(\frac{\d M}{\d t}\right)_{\rm dis} = -\frac{M}{t_{\rm dis}} = -\frac{M^{1-\gamma}}{t_0} ,
\end{equation}
which can be integrated for the mass evolution of the cluster due to dynamical dissolution.

The above formulation of the cluster mass evolution was extended to include stellar remnants, photometric cluster evolution, and a simple description of the MF in the {\tt SPACE} cluster models \citep{kruijssen08}. Stellar remnants were accounted for by assuming initial-final mass relations (similar to Sect.~\ref{sec:stev} of the present paper), while the photometric evolution was computed by integrating stellar isochrones from the Padova group \citep{bertelli94,girardi00}. The description of low-mass star depletion followed the simple model from \citet{lamers06} in which the minimum stellar mass of the MF increases with time.

The present study provides a new description of the evolution of the MF which is based on fundamental principles, {and does not depend on the above prescription for the total mass evolution}. In addition, the latest Padova models \citep{marigo08} are incorporated to calculate the photometric cluster evolution. These improvements update the {\tt SPACE} cluster models.

\section{The evolution of the stellar mass function} \label{sec:mf}
To describe the evolution of the MF, the effects of stellar evolution, stellar remnant production, and dynamical dissolution need to be included. While the focus of this paper lies with the effects of dissolution, a proper treatment of stellar evolution is essential. This is described first, before presenting a model for cluster dissolution.\footnote{The model presented in this paper is independent of the mass loss rate and of the form of the IMF $N_{\rm i}(m)$, but for explanatory purposes a \citet{kroupa01} IMF is adopted later on.}

\subsection{Stellar evolution} \label{sec:stev}
The influence of stellar evolution on the MF is twofold. First of all, the maximum stellar mass decreases, because at any time during cluster evolution the most massive stars reach the end of their lives. Secondly, the stellar remnants that are created upon the death of these massive stars constitute a part of the MF that can only be lost from the cluster by dynamical mechanisms.

The maximum stellar mass in the cluster as a function of its age is taken from the Padova 2008 isochrones \citep{marigo08} for metallicities in the range $Z=0.0001$---0.03. The stellar remnant masses $m_{\rm sr}$ are computed from their progenitor stellar mass $m$ using initial-final mass relations. Following \citet{kruijssen08}, for white dwarfs ($m<8~\msun$) the relation from \citet{kalirai08} is adopted:
\begin{equation}
\label{eq:wd}
  m_{\rm wd}=0.109m+0.394~\msun ,
\end{equation}
{which holds for all ages that are covered by the Padova isochrones}. For neutron stars ($8~\msun\leq m<30~\msun$) the relation from \citet{nomoto88} is used:
\begin{equation}
\label{eq:ns}
  m_{\rm ns}=0.03636(m-8~\msun)+1.02~\msun ,
\end{equation}
while for black holes ($m\geq 30~\msun$) a simple relation is assumed that is in acceptable agreement with theoretically predicted masses of stellar mass black holes \citep{fryer01}:
\begin{equation}
\label{eq:bh}
  m_{\rm bh}=0.06(m-30~\msun)+8.3~\msun .
\end{equation}
With these relations, the remnant MF is computed from conservation of numbers as
\begin{equation}
\label{eq:nsr}
  N_{\rm sr}(m_{\rm sr})=f_{\rm ret,sr}(M)N(m(m_{\rm sr}))\frac{\d m}{\d m_{\rm sr}} ,
\end{equation}
with ${\rm sr}=\{{\rm wd,ns,bh}\}$ denoting the appropriate remnant type, $N_{\rm sr}(m_{\rm sr})$ representing its MF, $f_{\rm ret,sr}(M)$ denoting the cluster mass-dependent fraction of these remnants that is retained after applying kick velocities, and $N(m(m_{\rm sr}))$ representing the progenitor MF. 

For a given velocity dispersion of remnants, the retention fraction of each remnant type depends on the local escape velocity $v_{\rm esc}$, which is related to the potential $\psi$ as $v_{\rm esc}=\sqrt{2\psi}$. Stellar remnants are {predominantly} produced in the cluster centre in the case of mass segregation, which is reached most rapidly for massive stars (see Sect.~\ref{sec:diss}). For a \citet[also see Eq.~\ref{eq:plummer}]{plummer11} potential this implies that upon remnant production $v_{\rm esc}=\sqrt{2GM/r_0}$, with $G$ the gravitational constant and $r_0$ the Plummer radius. Adopting a Maxwellian distribution of velocities that is truncated at $v_{\rm esc}$, it is straightforward to show that
\begin{equation}
\label{eq:fret}
  f_{\rm ret,sr}(x)=A\left[{\rm erf}\left(\frac{x}{\sqrt{2}}\right)-\sqrt{\frac{2}{\pi}}x{\rm e}^{-x^2/2}\right] ,
\end{equation}
where $A$ is a normalisation constant to account for the truncation of the velocity distribution and $x^2\equiv{2GM/r_0\sigma_{\rm sr}^2}$, with $\sigma_{\rm sr}^2={\sigma_0^2+\sigma_{\rm kick,sr}^2}$ denoting the total velocity dispersion of the produced remnant type, which arises from the central velocity dispersion in the cluster $\sigma_0^2={GM/3r_0}$ \citep[e.g.][]{heggiehut} and the velocity dispersion of the exerted kick $\sigma_{\rm kick}$. The normalisation constant then follows as $A={\rm erf}\sqrt{3}-2\sqrt{3/\pi}\exp{(-3)}$.

Typical values of the kick velocity dispersion $\sigma_{\rm kick,sr}$ are given in literature. White dwarf kicks have recently been proposed to be of order $\sigma_{\rm kick,wd}=4$~km~s$^{-1}$ \citep{davis08,fregeau09}. For neutron stars $\sigma_{\rm kick,ns}=100$~km~s$^{-1}$ is adopted, which is a somewhat conservative estimate with respect to theory, but it agrees reasonably well with observed neutron star numbers in globular clusters and represents a compromise between single star and binary channels \citep[for estimates of the retention fraction and discussions of the `neutron star retention problem' see][]{lyne94,drukier96,arzoumanian02,pfahl02}. Gravitational wave recoils are thought to exert black hole kicks of order $\sigma_{\rm kick,bh}=80$~km~s$^{-1}$ \citep{moody09}. This value depends on metallicity, but for simplicity I assume a single, typical value here.

\begin{figure}[t]
\resizebox{\hsize}{!}{\includegraphics{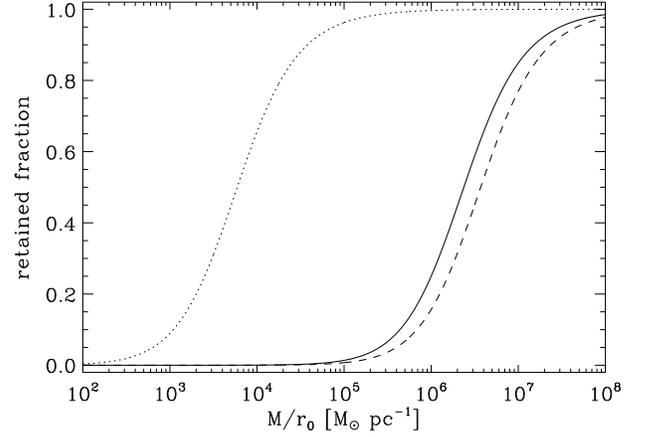}}
\caption[]{\label{fig:fret}
       Retention fraction of stellar remnants as a function of cluster mass per unit Plummer radius $M/r_0$, for black holes (solid), neutron stars (dashed) and white dwarfs (dotted).
                 }
\end{figure}
The retention fractions following from Eq.~\ref{eq:fret} are shown as a {function of cluster mass per unit Plummer radius in Fig.~\ref{fig:fret}. This quantity best reflects the retention fraction because $x^2\propto M/r_0$ in Eq.~\ref{eq:fret}.} Open clusters \citep[with initial masses $M_{\rm i}$ such that typically $M_{\rm i}/r_0<3\times10^4~\msun~{\rm pc}^{-1}$,][]{larsen04b} do not retain any neutron stars or black holes, while globular clusters \citep[$M_{\rm i}/r_0\sim3\times10^4$---$10^6~\msun~{\rm pc}^{-1}$,][]{harris96} retain 0.1---4\% of the neutron stars and 0.3---7\% of the black holes. These values are in excellent agreement with other studies \citep[e.g.][]{pfahl02,moody09}, but are still lower than the large observed number of neutron stars in a number of globular clusters (the aforementioned `retention problem').

\subsection{Dissolution and the evolution of the mass function} \label{sec:diss}
\begin{figure}[t]
\resizebox{\hsize}{!}{\includegraphics{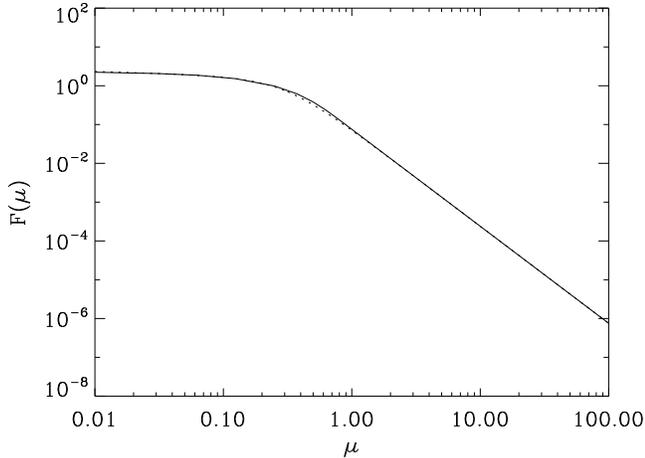}}
\caption[]{\label{fig:fhenon}
       H\'{e}non function $F(\mu)$, which is a measure for the ejection probability of a star of mass $m$ in a two-body interaction with mass ratio $\mu\equiv m/m'$. The dotted line shows the fit from Eq.~\ref{eq:fhenon}.
                 }
\end{figure}

Dissolution alters the shape of the stellar MF in star clusters due to the effects of two-body relaxation and energy equipartition. In a pioneering paper, \citet{henon69} derived the escape rate of stars of different masses from an isolated cluster. The cluster was represented by a \citet{plummer11} gravitational potential:
\begin{equation}
\label{eq:plummer}
  \psi(r)=\psi_0\left(1+\frac{r^2}{r^2_0}\right)^{-1/2} ,
\end{equation}
where $r_0$ denotes the Plummer radius setting the concentration of the cluster and $\psi_0\equiv GM/r_0$ represents the central potential, with $G$ the gravitational constant and $M$ the cluster mass. It was argued by \citet{henon60} that the only way for stars to escape such an isolated cluster is by a single, close encounter. The corresponding stellar mass-dependent escape rate was found to be \citep{henon69}:
\begin{equation}
\label{eq:henon}
  \frac{\d N(m)}{\d t}=-\frac{|E|^{3/2}N(m)}{GM^{9/2}}\int_0^{\infty}N(m')F\left(\frac{m}{m'}\right)m'^2\d m' ,
\end{equation}
with $N(m)$ the MF, $m$ the stellar mass, $E$ the total energy of the cluster, and $F(\mu)$ a function related to the ejection probability for a star of mass $m$ in a close encounter with a star of mass $m'$ and a corresponding mass ratio $\mu\equiv m/m'$.  {The expression in Eq.~\ref{eq:henon} is independent of the adopted IMF.} The function $F$ will be referred to as the `H\'{e}non function' and is shown in Fig.~\ref{fig:fhenon}. The original expression consists of several integrals that have to be solved numerically. In \citet{henon69}, a table is given for the H\'{e}non function, but it can also be fitted by:
\begin{equation}
\label{eq:fhenon}
 F(\mu)=\left(0.32+0.55\mu^{0.35}+13.26\mu^{2.5}\right)^{-1} .
\end{equation}
This approaches the power law $F(\mu)=0.075398\mu^{-5/2}$ for $\mu>1$, as was derived explicitly by \citet{henon69}.

The total mass loss rate corresponding to Eq.~\ref{eq:henon} conflicts with $N$-body simulations \citep[as was already noted by][]{wielen71} because only ejections by single, close encounters are {included}. This restriction implies that the disruption timescale $t_{\rm dis}$ is proportional to the crossing time ($t_{\rm dis,H\acute{e}non}\propto t_{\rm cr}$), while $N$-body simulations show that it scales with a combination of the half-mass relaxation time and the crossing time ($t_{\rm dis}\propto t_{\rm rh}^{0.75}t_{\rm cr}^{0.25}$) due to two-body relaxation, i.e. the repeated effect of soft encounters \citep[e.g.][]{baumgardt03}. Nonetheless, the escape rate from \citet{henon69} does accurately describe what happens if two stars interact and can therefore be used as a starting point for a more complete description of the evolution of the MF. For that purpose, it is convenient to scale Eq.~\ref{eq:henon} to the mass loss rate found in $N$-body simulations and only use the relative or `differential' stellar mass dependence from \citet{henon69}. This is allowed if the ratio $t_{\rm dis}/t_{\rm dis,H\acute{e}non}=(t_{\rm rh}/t_{\rm cr})^{0.75}$ only depends on global cluster properties. It is straightforward to show \citep[e.g.][]{spitzer87,heggiehut} that indeed this is the case as $t_{\rm rh}/t_{\rm cr}\propto N/\ln{\Lambda}$ with $\ln{\Lambda}$ the Coulomb logarithm. As such, one can write
\begin{equation}
\label{eq:dndt}
  \frac{\d N(m)}{\d t}=\left(\frac{\d M}{\d t}\right)_{\rm dis}\chi(m) ,
\end{equation}
with $(\d M/\d t)_{\rm dis}$ the mass loss rate from Eq.~\ref{eq:dmdt} \citep{lamers05} and $\chi(m)$ the stellar mass-dependent escape rate per unit mass loss rate. The quantity $\chi(m)$ is completely independent of the prescription for the total mass evolution. In order to derive $\chi(m)$, I start from Eq.~\ref{eq:henon} and express $\chi(m)$ as
\begin{eqnarray}
\label{eq:chi}
  \lefteqn{\chi(m)=}\\
  \nonumber\lefteqn{\frac{N(m)\int_0^{\infty}N(m')F(m/m')\lambda(m,m')m'^2\d m'}{\int_0^{\infty}m''N(m'')\int_0^{\infty}N(m')F(m''/m')\lambda(m'',m')m'^2\d m'\d m''},}
\end{eqnarray}
where $\lambda(m,m')$ represents a correction factor to account for additional physics (see below). The numerator reflects the escape rate, while the denominator is proportional to the mass loss rate.

For mathematical simplicity\footnote{And because this is the only way to obtain an analytical solution as in Eq.~\ref{eq:henon}.} \citet{henon69} made the following assumptions in the derivation of Eq.~\ref{eq:henon}.
\begin{itemize}
\item[(1)]
The cluster exists in isolation and the tidal field is neglected. Therefore, ejection can only occur by a single, close encounter and the repeated effect of soft encounters (two-body relaxation) is not accounted for. This underestimates the escape rate of massive stars.
\item[(2)]
The {distribution of stars is independent of stellar mass, i.e.} mass segregation is not included. Depending on the balance between their encounter rate and their proximity to the escape energy, this over- or underestimates the escape rate of low-mass stars from \citet{henon69}. Considering the results from \citet{baumgardt03}, the latter seems to be the case.
\end{itemize}
The remainder of this section concerns the derivation of the factor $\lambda(m,m')$ in Eq.~\ref{eq:chi} that corrects for the above assumptions.

Let us assume that the distribution of stars over radius and velocity space is initially independent of their mass. This implies that mass segregation is dynamically created and not primordial, which is discussed in Sect.~\ref{sec:disc}. For such an initial distribution, the separation from the escape energy $\Delta E$ is independent of mass. As the cluster evolves, energy equipartition is reached between the stars and the radius, velocity and proximity to the escape energy become a function of stellar mass. I first consider this effect on the escape rate before including the timescale on which two-body relaxation occurs for different stellar masses. Please note that the formulation of Eq.~\ref{eq:chi} with $\lambda(m,m')$ appearing in the numerator and the denominator implies that only the proportionality of $\lambda(m,m')$ is important. Its exact value is determined by constants that drop out when substituting in Eq.~\ref{eq:chi}.

It is intuitive to express the dependence of the escape rate on the energy needed for escape as $\d N(m)/\d t\propto [\Delta E(m)]^{-1}$. The energy that is required for escape $\Delta E$ is related to the position and velocity of the star.\footnote{The energy difference $\Delta E$ that is discussed here concerns the energy that needs to be added to reach the escape energy. As such, it differs from the separation from the escape energy in \citet{fukushige00} and \citet{baumgardt01}, who are considering the {\it excess} energy of stars and its relation to the escape time, resulting in the aformentioned relation $t_{\rm dis}\propto t_{\rm rh}^{0.75}t_{\rm cr}^{0.25}$.} For the potential in Eq.~\ref{eq:plummer} it is given by
\begin{equation}
\label{eq:eesc}
  \Delta E(r,v)=\psi(r)-\frac{v^2}{2}=\psi_0\left(1+\frac{r^2}{r^2_0}\right)^{-1/2}-\frac{v^2}{2} ,
\end{equation}
with $r$ and $v$ the radial position and velocity of the star, and $v_{\rm esc}\equiv\sqrt{2\psi(r)}$ its escape velocity. If the cluster is in `perfect' energy equipartition and correspondingly perfect mass segregation, the radius and velocity become a {monotonous} function of stellar mass \citep[Ch. 16]{heggiehut}. Mass segregation is strongest in the cluster centre, which for a \citet{plummer11} potential can be approximated with a harmonic potential $\psi\propto r^2$. For a cluster in a tidal field the potential is truncated, and the harmonic approximation serves as a crude but reasonable approximation for the entire cluster \citep[Ch. 16]{heggiehut}. Energy equipartition yields
\begin{equation}
\label{eq:veq}
  v^2(m)=\langle v\rangle^2\frac{\langle m\rangle}{m} ,
\end{equation}
with $\langle v\rangle^2\propto\psi_0$ the mean speed of all stars squared and $\langle m\rangle$ the mean stellar mass. For the harmonic potential, this translates to a similar relation for the radial position:
\begin{equation}
\label{eq:req}
  r(m)=r_0\sqrt{\frac{\langle m\rangle}{m}} ,
\end{equation}
where $r_0$ represents the typical radius of the system, in this case the Plummer radius. This relation assumes that there is no particular stellar mass which dominates the mass spectrum. The decrease of radial position with stellar mass implied by Eq.~\ref{eq:req} is a direct consequence of the energy loss endured by massive stars\footnote{And the energy {\it gain} experienced by low-mass stars.} as the system evolves towards energy equipartition. Substituting Eqs.~\ref{eq:veq} and~\ref{eq:req} into Eq.~\ref{eq:eesc} and dividing out the proportionality $\langle v\rangle^2\propto\psi_0$ gives an expression for $\Delta E(m)$:
\begin{equation}
\label{eq:eescm}
  \Delta E(m)=\left(1+\frac{\langle m\rangle}{m}\right)^{-1/2}-c_1\frac{\langle m\rangle}{m} ,
\end{equation}
with $c_1=\langle v\rangle^2/2\psi_0$ denoting the ratio of the mean speed squared to the central escape velocity squared. This constant mainly depends on the degree of mass segregation. Consequently, it will depend on the IMF. By comparing the models to the $N$-body simulations with a mass spectrum by \citet{baumgardt03} the value is constrained to $c_1=0.020$ for a Kroupa IMF, using \citet{king66} potentials with King parameter $W_0=5$---7 (see Sect.~\ref{sec:comp}). For reference, an unevolved \citet{plummer11} potential has $\langle v\rangle^2/2\psi_0=3\pi/64=0.147$. 

{By comparing the models to $N$-body simulations (provided by M. Gieles, private communication) with different IMF power law slopes $\alpha$ and a ratio between the maximum and minimum mass of 10, the approximate relation $\log{c_1}\approx\alpha-3.76$ is found\footnote{This prescription for $c_1$ implies that the condition for the stars in the cluster to be physically bound $\Delta E(m)>0$ is satisfied for all $\alpha<3.63$.} for a MF $n_{\rm s}\propto m^{-\alpha}$. Fitting the Kroupa IMF with a single power law in the mass range 0.08---$15~\msun$ \citep[as used by][]{baumgardt03} yields $\alpha=2.06$, resulting in $c_1=0.020$ as mentioned earlier.\footnote{Nonetheless, the relation for $c_1$ should be expected to exhibit some variation for different mass ranges.} The comparison with $N$-body simulations also showed that a single value of $c_1$ suffices to determine the MF evolution, even though it does not remain constant over the full cluster lifetime.}

Because $\d N(m)/\d t\propto [\Delta E(m)]^{-1}$, Eq.~\ref{eq:eescm} indicates that the escape rate of low-mass stars is increased if a cluster is in complete energy equipartition. However, the timescale on which two-body relaxation occurs between different stellar masses has not yet been considered. For a cluster starting with a {stellar mass-independent} distribution of radial positions and velocities, the equipartition timescale $t_{\rm e}$ scales as
\begin{equation}
\label{eq:teq}
  t_{\rm e}(m,m')\propto m^{-1}m'^{-1} ,
\end{equation}
for equipartition between stars of masses $m$ and $m'$ \citep{heggiehut}. This is a modified version of the relaxation timescale, which shows a very similar proportionality ($t_{\rm r}\propto m^{-2}$). It illustrates that two-body relaxation occurs on a shorter timescale for massive stars than for low-mass stars, increasing their escape rate $\d N(m)/\d t\propto t_{\rm e}^{-1}$.

The correction factor for the escape rate $\lambda(m,m')$ that appears in the integrals of Eq.~\ref{eq:chi} now follows from Eqs.~\ref{eq:eescm} and~\ref{eq:teq} as
\begin{eqnarray}
\label{eq:lambda}
  \nonumber \lambda(m,m')&=&t_{\rm e}^{-1}(m,m')[\Delta E(m)]^{-1} \\
                   &=&mm'\left[\left(1+\frac{\langle m\rangle}{m}\right)^{-1/2}-c_1\frac{\langle m\rangle}{m}\right]^{-1} .
\end{eqnarray}
It was mentioned before that the proportionalities of $\Delta E(m)$ and $t_{\rm e}(m,m')$ rather than their exact values suffice for the computation of $\lambda(m,m')$ due to the renormalisation of the total mass loss rate that appears in Eq.~\ref{eq:chi}: {\it only the stellar mass-dependence is important}. 

The influence of the tidal field is now included in two ways. First of all, the ejection of stars no longer occurs by a single, close encounter but arises due to two-body relaxation on the equipartition timescale, representing the repeated effect of soft encounters. Secondly, the above derivation of the separation from the escape energy assumes a potential which approximates tidally limited clusters. As a result, the escape rate of massive stars is increased with respect to clusters in the model of \citet{henon69}, which was derived for an isolated cluster. On the other hand, the effect of mass segregation is included by introducing a stellar mass-dependence for the energy needed by stars to reach the escape velocity. Low-mass stars are closer to the tidal radius than massive stars, leading to a lower energy that is needed for escape and an increased escape rate. It depends on the shape of the MF which mechanism dominates.

The evolution of the MF of various cluster components is obtained from Eqs.~\ref{eq:dndt},~\ref{eq:chi} and~\ref{eq:lambda} by writing
\begin{equation}
\label{eq:dndtall}
  \frac{\d\log{N_{\rm comp}(m)}}{\d t} = \frac{\d\log{N(m)}}{\d t} ,
\end{equation}
where the MFs of stars, white dwarfs, neutron stars and black holes are represented by $N_{\rm comp}(m)$, with ${\rm comp}=\{{\rm s, wd, ns, bh}\}$. The overall cluster evolution is computed by combining the results of this section with the prescription for stellar evolution from Sect.~\ref{sec:stev}.

{If stellar evolution is included, the resulting mass loss causes an expansion of the cluster, during which stars are lost independently of their masses. This delays the onset of mass segregation and the stellar mass-dependent mass loss that is described above.} The moment of transition to stellar mass-dependent mass loss can be characterised by a certain fraction of the initial cluster mass that has been lost by dissolution $f_{\rm diss}\equiv M_{\rm diss}/M_{\rm i}$. It is assumed that the fraction $f_{\rm smd}$ of the mass loss for which the ejection rate depends on the stellar mass grows exponentially\footnote{This form assumes that the increase of the fraction of the mass loss that is stellar mass-dependent scales with the total dynamical mass loss, which is a compromise between a step function and a linear increase.} between 0 and 1 as
\begin{equation}
\label{eq:fsmd}
  f_{\rm smd}=C\left({\rm e}^{f_{\rm diss}/f_{\rm diss,seg}}-1\right) ,
\end{equation}
where the subscript `smd' denotes `stellar mass-dependent', $f_{\rm diss,seg}\equiv M_{\rm diss,seg}/M_{\rm i}$ is the fraction of the initial mass that has been lost by dissolution at which mass segregation is reached, and $C=({\rm e}-1)^{-1}$ is a constant to normalise $f_{\rm smd}=1$ at the reference value $f_{\rm diss}=f_{\rm diss,seg}$. For $f_{\rm diss}>f_{\rm diss,seg}$, per definition $f_{\rm smd}=1$, indicating that all mass loss is stellar mass-dependent. The timescale $t_{\rm seg}$ on which mass segregation is reached and the transition to stellar mass-dependent mass loss is completed is proportional to the initial half-mass relaxation time ($t_{\rm seg}\propto t_{\rm rh,i}$). It has been shown in several studies that for Roche lobe-filling clusters the disruption timescale $t_{\rm dis}\propto t_{\rm rh,i}^{0.75}t_{\rm cr}^{0.25}$ \citep{vesperini97b,baumgardt03,gieles08}, implying that $t_{\rm seg}/t_{\rm dis}\propto t_{\rm dis}^{0.33}$. The expression for $t_{\rm dis}$ in Eq.~\ref{eq:tdis} then leads to $t_{\rm seg}/t_{\rm dis}\propto t_0^{0.33}M_{\rm i}^{0.33\gamma}$. Assuming that the cluster mass evolution is close to linear, the first-order relation $f_{\rm diss,seg}\propto t_{\rm seg}/t_{\rm dis}$ is obtained, implying
\begin{equation}
\label{eq:tseg}
  f_{\rm diss,seg}=c_2\left(\frac{t_0}{t_0^\odot}\right)^{0.33}\left(\frac{M_{\rm i}}{10^4~\msun}\right)^{0.21} ,
\end{equation}
for a King parameter of $W_0=5$, with the dissolution timescale at the solar galactocentric radius $t_0^\odot=21.3$~Myr. For a King parameter of $W_0=7$, the exponent of the initial cluster mass $M_{\rm i}$ becomes 0.23 and $t_0^\odot=10.7$~Myr \citep{kruijssen09}. In this relation, $c_2$ represents a constant that is fixed by comparing the model to the results of $N$-body simulations from \citet{baumgardt03}, giving $c_2=0.25$ for $W_0=5$ and $c_2=0.15$ for $W_0=7$ (see Sect.~\ref{sec:comp}). The variation with King parameter arises because two-body relaxation is faster for more concentrated clusters. {If stellar evolution were neglected, at all ages $c_2=0$ and $f_{\rm smd}=1$.}

\begin{figure}[t]
\resizebox{\hsize}{!}{\includegraphics{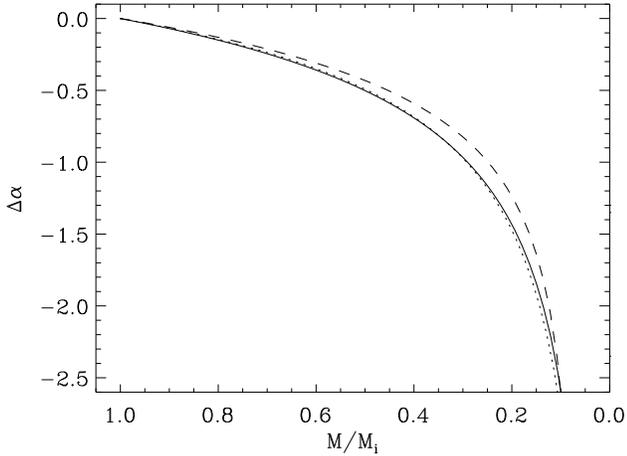}}
\caption[]{\label{fig:alpha}
       {MF slope change $\Delta\alpha$ in the range $m=0.1$---$0.5~\msun$ versus the remaining mass fraction for a Kroupa IMF (solid), Salpeter IMF (dotted), and a power law IMF with $\alpha=1.35$ (dashed). In all cases, the IMF mass range is $m=0.1$---$1~\msun$. The displayed relation is valid if stellar evolution is excluded.}
                 }
\end{figure}
{The modeled MF slope change $\Delta\alpha$ in the mass range $m=0.1$---$0.5~\msun$ is shown in Fig.~\ref{fig:alpha} for different IMFs covering $m=0.1$---$1~\msun$. Evidently, $\Delta\alpha$ is a function of the remaining mass fraction and is insensitive to the slope of the IMF, as long as that the ratio between the maximum and minimum mass is kept fixed and stellar evolution is excluded. This is an interesting observation in view of the MF evolution of globular clusters, in which $m\approx 0.1$---$1~\msun$ and stellar evolution only plays a minor role. Figure~\ref{fig:alpha} shows that the slope of the MF in globular clusters could be a possible indicator for the mass fraction that has been lost due to dissolution, provided that the IMF does not vary and the remnant retention fractions were not substantially dissimilar during the early evolution of different globular clusters (see Sect.~\ref{sec:remn} and Fig.~\ref{fig:alphatdis}).}

\begin{figure}[t]
\resizebox{\hsize}{!}{\includegraphics{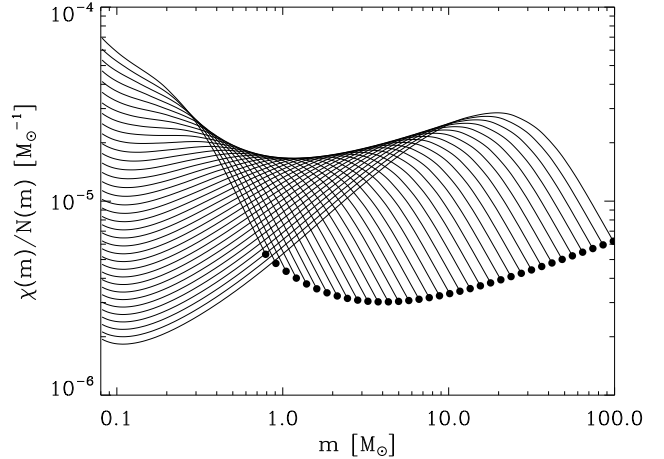}}
\caption[]{\label{fig:chinorm}
       Relative ejection rate $\chi(m)/N(m)$ as a function of stellar mass, shown for a Kroupa MF with different maximum masses. The end point of each curve (dot) marks its maximum mass. The quantity $\chi(m)/N(m)\equiv (\d\log{N(m)}/\d t)/(\d M/\d t)$ represents the ejection rate per unit mass loss rate normalised to the number of stars at each mass (also see Eq.~\ref{eq:chi}).
                 }
\end{figure}
For the particular example of a Kroupa MF that is truncated at different maximum masses $m_{\rm max}$, the relative escape rate per unit mass loss rate $\chi(m)/N(m)$ (see Eqs.~\ref{eq:dndt} and~\ref{eq:chi}) is shown in Fig.~\ref{fig:chinorm}. This quantity is proportional to $\d\log{N(m)}/\d t$ and reflects the probability that a star of a certain mass is ejected. Figure~\ref{fig:chinorm} illustrates that the mass of the highest relative ejection rate is related to the maximum mass of the MF. The peak occurs at intermediate masses if the MF is truncated at a high mass. This implies that there is a typical mass where the stars are not too far from the escape energy {\it and} have an equipartition timescale with the massive stars that is short enough to eject them efficiently. This `sweet spot' depends on the maximum mass of the MF. If the MF is truncated at an intermediate mass, the combination of quick two-body relaxation and proximity to the escape energy favours the ejection rate of stars at the lowest masses. 

\begin{figure}[t]
\resizebox{\hsize}{!}{\includegraphics{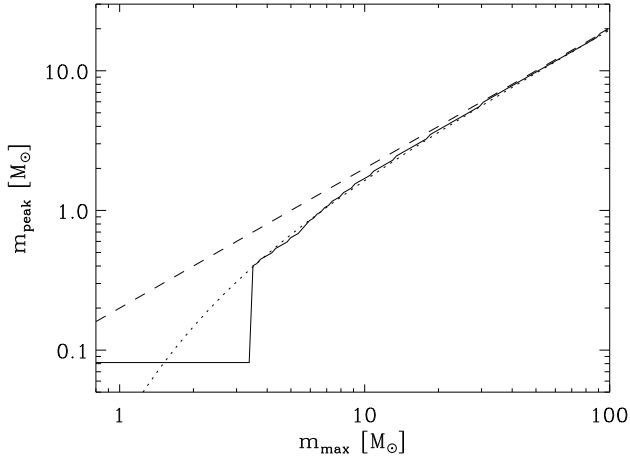}}
\caption[]{\label{fig:maxchinorm}
       Mass of the highest relative ejection rate $m_{\rm peak}$ as a function of the maximum stellar mass of the MF $m_{\rm max}$ (solid line). The dashed line represents the relation $m_{\rm peak}=0.2m_{\rm max}$, while the dotted line describes an eyeball fit for masses $m_{\rm max}>3~\msun$ and includes an exponential truncation at the low-mass end (see Eq.~\ref{eq:mpeak}).
                 }
\end{figure}
The maximum stellar mass at which the transition from `sweet spot'-depletion to low-mass star depletion happens, is determined by the proximity of the low-mass stars to the escape energy. In Fig.~\ref{fig:maxchinorm}, the mass of the peak relative ejection rate is shown as a function of the maximum stellar mass. At low truncation masses, the peak occurs at the minimum mass, indicating strong low-mass star depletion. Around $m_{\rm max}\sim 3~\msun$, the relative ejection rate at $m_{\rm peak}\sim 0.4~\msun$ becomes larger than its value at the lowest masses, which causes a jump in Fig.~\ref{fig:maxchinorm}. For even higher values of $m_{\rm max}$, the peak relative ejection rate typically occurs at 15---20\% of the maximum mass, approximately following the relation
\begin{equation}
\label{eq:mpeak}
  m_{\rm peak}=0.2m_{\rm max}{\rm e}^{-2M_\odot/m_{\rm max}} .
\end{equation}

Even though its quantitative properties only hold for a Kroupa MF, the variation of the relative escape rate with the maximum mass of the MF has several implications for star cluster evolution. The change of $m_{\rm max}$ in Figs.~\ref{fig:chinorm} and~\ref{fig:maxchinorm} can be interpreted as an example of what happens when stellar evolution removes the most massive stars in the cluster, provided that the remnants are all ejected by their kick velocities. If dynamical evolution does not affect the shape of the MF too much before $m_{\rm max}(t)\sim 3~\msun$, or $t\sim 400$~Myr, the subsequent evolution of the MF will be dominated by low-mass star depletion. If substantial dissolution occurs earlier on, it is dominated by the `sweet spot' depletion of intermediate masses. Only the retention of massive stellar remnants will make the evolution of the MF deviate from these basic estimates, because remnant retention can provide a fixed maximum (remnant) mass of the MF. This is treated in more detail in Sect.~\ref{sec:results}.

\section{Comparison to $N$-body simulations} \label{sec:comp}
\begin{figure*}[t]
\resizebox{\hsize}{!}{\includegraphics{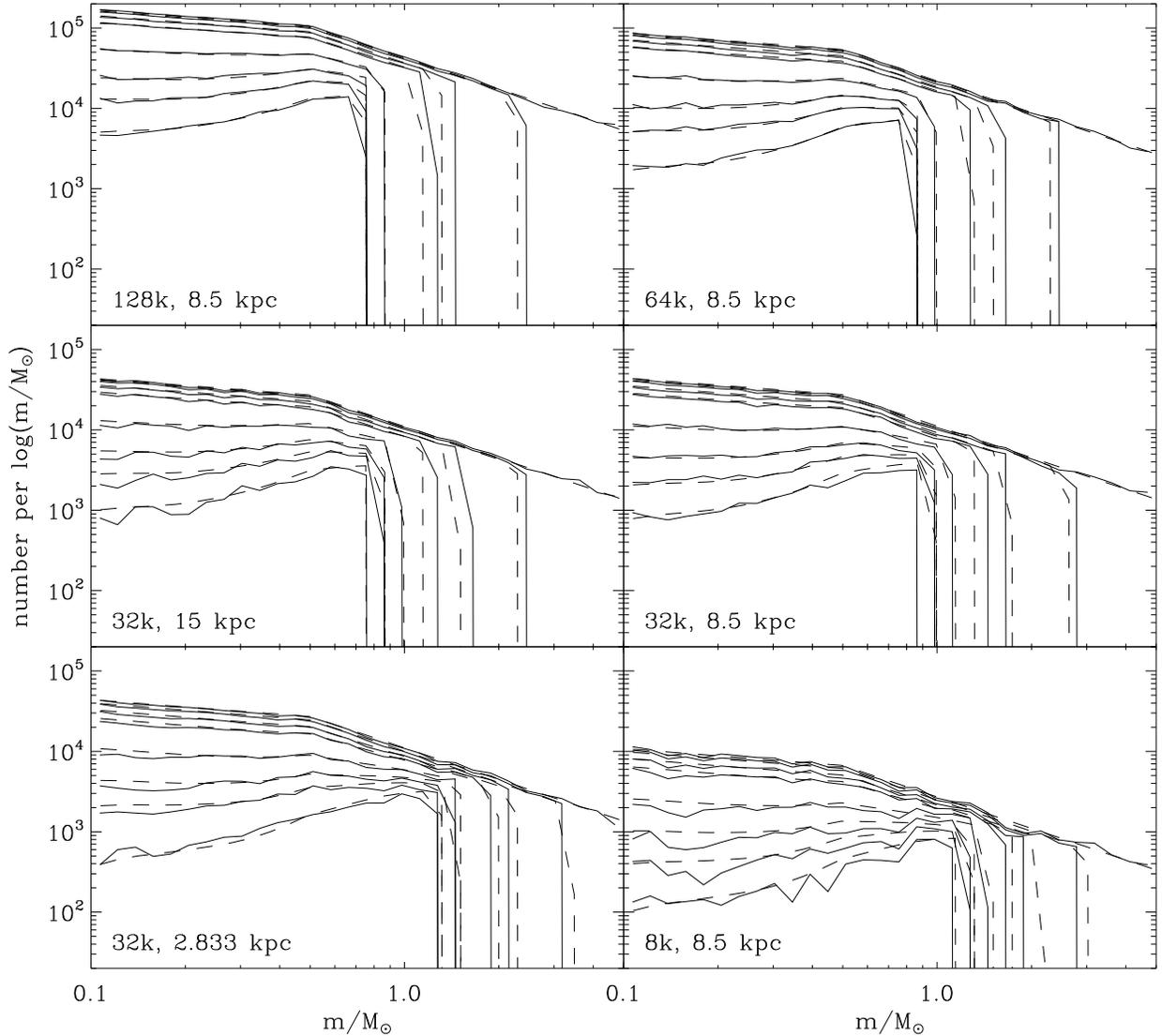}}
\caption[]{\label{fig:comp}
       Comparison of the evolution of the stellar MF from the models (dashed) to the $N$-body runs from \citet[solid]{baumgardt03} for the exact same boundary conditions. The initial number of particles and the galactocentric radius are indicated in the bottom-left corner of each panel. From top to bottom, the subsequent MFs in each panel are shown for the times at which the remaining cluster mass fraction equals $M/M_{\rm i}=\{1,0.75,0.6,0.5,0.3,0.2,0.15,0.1\}$.
                 }
\end{figure*}
The model described in Sect.~\ref{sec:mf} can be easily verified by running it for the exact same boundary conditions as the $N$-body simulations\footnote{These were performed using {\tt NBODY4} \citep{aarseth99}.} by \citet{baumgardt03} and comparing the results. {They conducted simulations of Roche lobe-filling clusters between 8k and 128k particles, which were evolved in the Galactic tidal field at galactocentric radii in the range 2.833---15~kpc.} The boundary conditions for the $N$-body runs of \citet{baumgardt03} differ from those described in Sect.~\ref{sec:mf} by neglecting kick velocities and defining the Kroupa stellar IMF between 0.1 and 15~$\msun$, thereby excluding black holes. For this particular comparison, the same IMF, stellar evolution prescription, and initial-final mass relation for stellar remnants are used in the model that is presented in this paper. 

In Fig.~\ref{fig:comp}, the modeled evolution of the (luminous) stellar MF is compared to the $N$-body runs with King parameter $W_0=5$ for a range of cluster masses and total disruption times. As time progresses, the maximum stellar mass decreases due to stellar evolution and the MF is lowered due to the dynamical dissolution of the star cluster. The slope of the MF changes due to the preferential ejection of low-mass stars, which have energies closer to their escape energies, even to the extent that it dominates over their relatively slow two-body relaxation. For both the models and the $N$-body simulations, the MF develops a slight bend at $m\sim 0.3~\msun$ when approaching total disruption. The bend arises as an optimum between on the one hand high energies but slow relaxation for the lowest stellar masses, and on the other hand quick relaxation but low energies for the highest stellar masses (see the discussion at the end of Sect.~\ref{sec:mf}).

In all cases, the resemblance of the models and the $N$-body simulations is striking. The models reproduce all key aspects of the $N$-body runs, such as the amount of low-mass star depletion, the changing slope at $m\sim 0.3~\msun$ for clusters close to dissolution, the survival of the Kroupa bend at $m=0.5~\msun$, and the dependence of the low-mass depletion on the total lifetime of the cluster (compare the three 32k runs). The only difference occurs at the high-mass end of the MF, where the maximum stellar masses do not match at young ages. This is due to a minor dissimilarity of the total mass evolution \citep[also see][]{lamers05,kruijssen08}. Because the maximum stellar mass only depends on the age of the cluster, this causes a difference in maximum stellar mass when showing the MFs at fixed remaining cluster mass fractions. The contrast is clearest at young ages, since there the maximum stellar mass most rapidly decreases.

\begin{figure}[t]
\resizebox{\hsize}{!}{\includegraphics{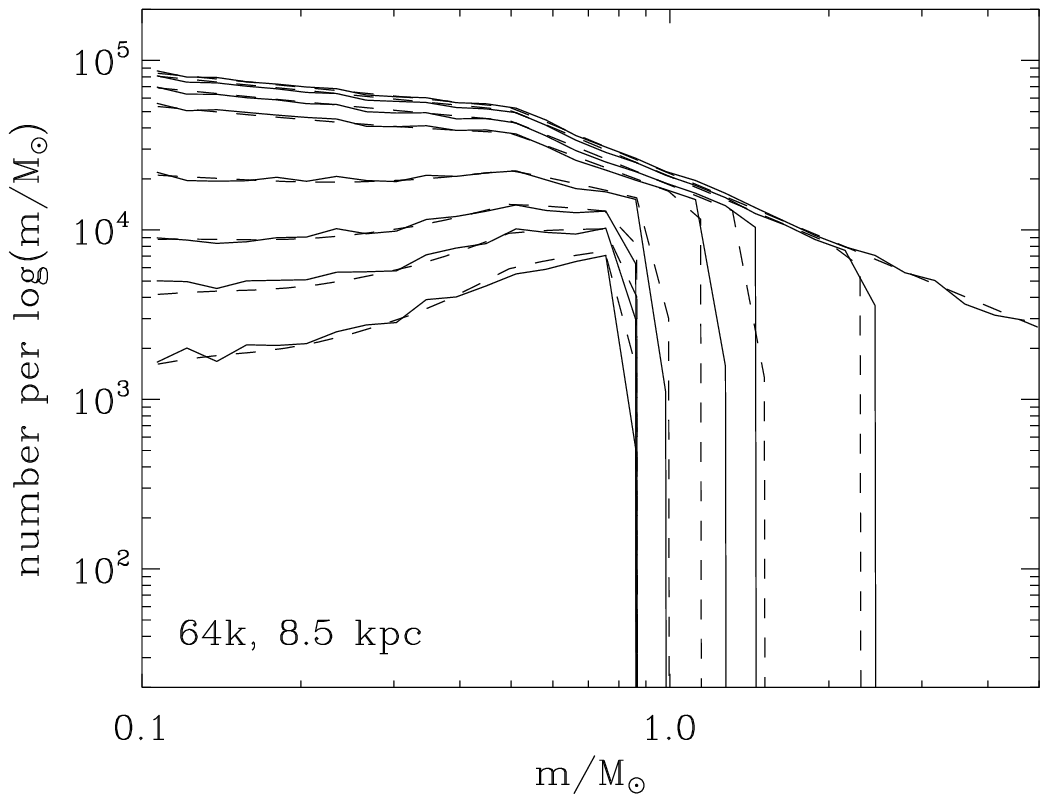}}
\caption[]{\label{fig:comp2}
       Comparison of the evolution of the stellar MF from the models (dashed) to the $N$-body run from \citet[solid]{baumgardt03} with $W_0=7$ for the exact same boundary conditions. From top to bottom, the subsequent MFs are shown for the times at which the remaining cluster mass fraction equals $M/M_{\rm i}=\{1,0.75,0.6,0.5,0.3,0.2,0.15,0.1\}$.
                 }
\end{figure}

In the description of the model in Sect.~\ref{sec:mf}, two constants have been determined from the $N$-body simulations by \citet{baumgardt03}. These constants are the ratio of the mean speed squared to the central escape velocity squared ($c_1$, see Eq.~\ref{eq:eescm}) and the proportionality constant for the relation marking the transition to stellar mass-dependent mass loss ($c_2$, see Eq.~\ref{eq:tseg}). As mentioned in Sect.~\ref{sec:mf}, for a Kroupa IMF and King parameter $W_0=5$ one obtains $c_1=0.020$ and $c_2=0.25$. To illustrate the robustness of the models, in Fig.~\ref{fig:comp2} they are compared to a 64k $N$-body run with $W_0=7$. For such a cluster with a higher concentration, the early mass segregation implies $c_2=0.15$. Again, the model and the simulation are in excellent agreement. 

\begin{figure}[t]
\resizebox{\hsize}{!}{\includegraphics{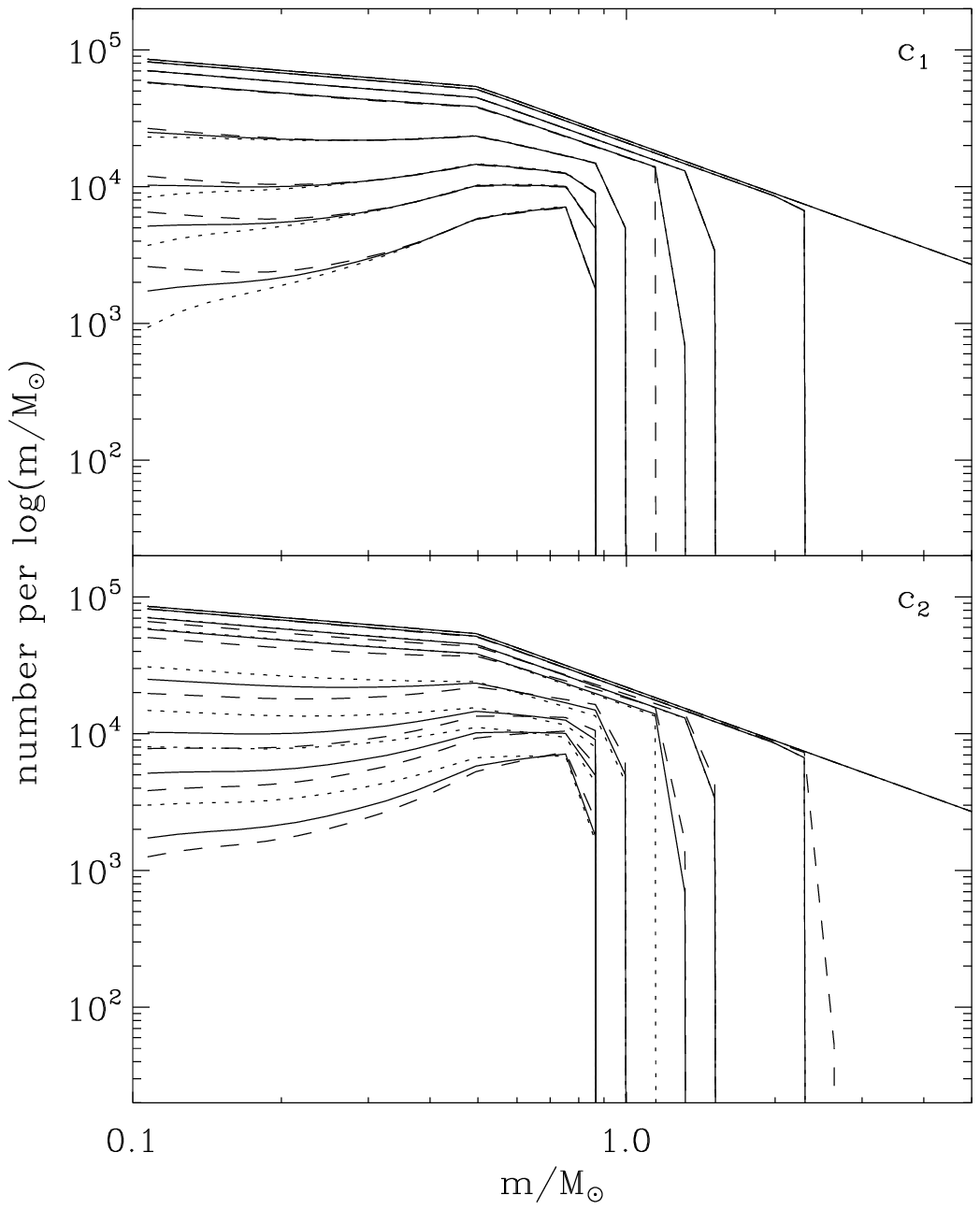}}
\caption[]{\label{fig:constants}
       Influence of the constants $c_1$ and $c_2$ on the evolution of the stellar MF. From top to bottom, the subsequent MFs in each panel are shown for the times at which the remaining cluster mass fraction equals $M/M_{\rm i}=\{1,0.75,0.6,0.5,0.3,0.2,0.15,0.1\}$. {\it Top panel}: the values $c_1=\{0.010,{\bf0.020},0.030\}$ are represented by dashed, solid and dotted lines, respectively. {\it Bottom panel}: the values $c_2=\{0,{\bf0.25},0.40\}$ are represented by dashed, solid and dotted lines, respectively. For both $c_1$ and $c_2$, the second (boldfaced) value is the one obtained from the comparison to the $N$-body simulations with $W_0=5$ in Fig.~\ref{fig:comp}.
                 }
\end{figure}

The dependence of the MF evolution on both constants is considered in Fig.~\ref{fig:constants}. For $c_1$, the dependence of the evolution of the MF on its value is shown in the upper panel of Fig.~\ref{fig:constants}, while for $c_2$ it is shown in the bottom panel of Fig.~\ref{fig:constants}. Both panels show the evolution of the MF for the 64k cluster in Fig.~\ref{fig:comp} for different values of $c_1$ and $c_2$.

The ratio of the mean speed squared to the central escape velocity squared $c_1$ affects the ejection probability of the stars with the lowest masses. Because these stars are closest to their escape energies in a mass-segregated cluster, they are most strongly influenced by the value of $c_1$. For higher $c_1$, the MF gets more depleted in low-mass stars due to their closer proximity to the escape energy, while for lower $c_1$ more low-mass stars are retained as the balance between close proximity to the escape energy and slow relaxation shifts to the latter.

The proportionality constant for the transition to stellar mass-dependent dissolution $c_2$ in Eq.~\ref{eq:tseg} affects the MF as a whole. For lower $c_2$, the transition occurs earlier and more low-mass stars are lost, while for higher $c_2$ the onset of the depletion is delayed and the slope of the MF remains closer to its initial value. If one were to assume a constant $f_{\rm diss,seg}$, which is contrary to the adopted relation with cluster mass in Eq.~\ref{eq:tseg}, this would therefore yield a stellar MF in massive clusters that is underpopulated in low-mass stars, and a MF in low-mass clusters that is overabundant in low-mass stars.

\section{Star cluster evolution} \label{sec:results}
In this section, the described model is applied to compute the evolution of clusters for a variety of boundary conditions. The stellar content as well as integrated photometry are addressed, using the boundary conditions from Sect.~\ref{sec:mf} instead of those that were adopted to compare the model to $N$-body simulations in Sect.~\ref{sec:comp}. The most important differences are the mass range of the IMF, the inclusion of remnant kick velocities, and the initial-final mass relation.

The model that will be referred to as the `standard model' uses a metallicity $Z=0.004$ {(which is typical of globular clusters)}, a King parameter\footnote{For $W_0=5$, or $\gamma=0.62$, the results vary only marginally.} of $W_0=7$ (corresponding to $\gamma=0.7$ in Eq.~\ref{eq:tdis}), a dissolution timescale parameter $t_0=1$~Myr, and a Kroupa IMF between $m=0.08~\msun$ and the maximum stellar mass given by the Padova isochrones at $\log{t}=6.6$, which is typically $m\sim 70~\msun$. For the computation of the retained remnant fraction (see Eq.~\ref{eq:fret}), the Plummer radius $r_0$ is related to the half-mass radius $r_{\rm h}$ as $r_{\rm h}=1.3r_0$. The half-mass radius is assumed to remain constant during the cluster lifetime \citep[e.g.][]{aarseth98}. For the relation between $r_{\rm h}$ and initial cluster mass $M_{\rm i}$ the expression from \citet{larsen04b} is adopted:
\begin{equation}
\label{eq:rh}
r_{\rm h}=3.75~{\rm pc}\left(\frac{M_{\rm i}}{10^4~\msun}\right)^{0.1} .
\end{equation}
The models that are used in this section are computed from $10^7$~yr to $1.65\times 10^{10}$~yr (the maximum age of the Padova isochrones) for initial masses between $10^2~\msun$ and $10^7~\msun$, spaced by 0.25~dex intervals.

\subsection{The influence of the disruption time} \label{sec:tdis}
The disruption time of a cluster affects the evolution of the MF and of the integrated photometric properties. To assess the influence of the disruption time on cluster evolution, clusters with low and high remnant retention fractions should be treated separately, because the presence of massive remnants also has a pronounced effect on the results (see Sect.~\ref{sec:remn}). As shown in Fig.~\ref{fig:fret}, for a given kick velocity dispersion the remnant retention fraction is set by the cluster mass. This means that the division between low and high remnant retention fractions can be made by making a cut in initial cluster mass. 

\begin{figure}[t]
\resizebox{\hsize}{!}{\includegraphics{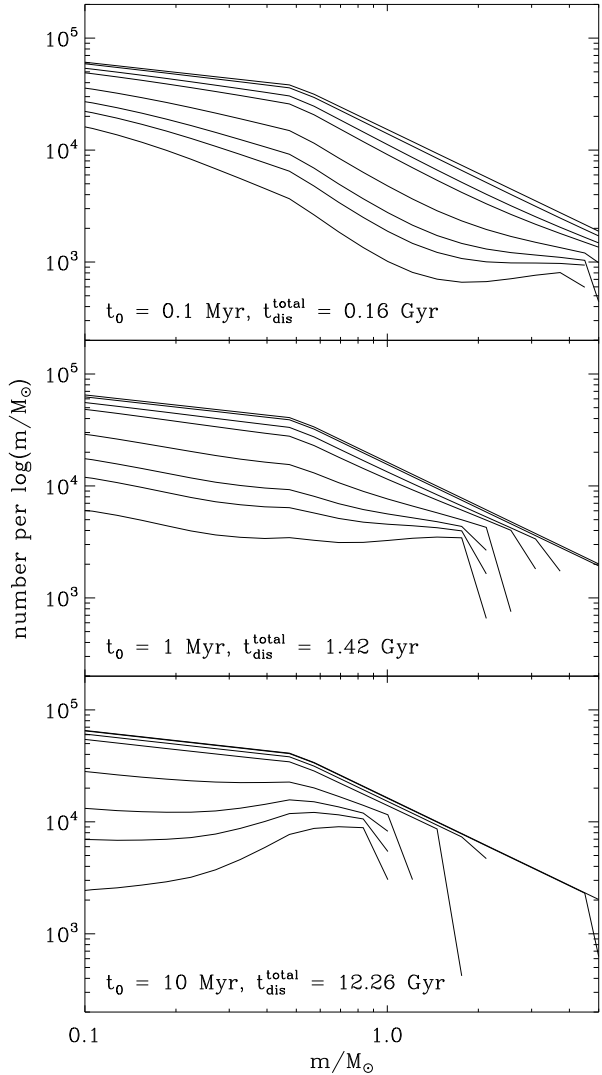}}
\caption[]{\label{fig:mf1}
       Influence of the disruption time on the evolution of the stellar MF for a cluster with a low remnant retention fraction ($\log{(M_{\rm i}/\msun)}=4.5$). From top to bottom, the subsequent MFs in each panel are shown for the times at which the remaining cluster mass fraction equals $M/M_{\rm i}=\{1,0.75,0.6,0.5,0.3,0.2,0.15,0.1\}$.
                 }
\end{figure}
In Fig.~\ref{fig:mf1}, the impact of the disruption time on the evolution of the MF is shown for a cluster with initial mass $\log{(M_{\rm i}/\msun)}=4.5$, representing the evolution for low remnant retention fractions.\footnote{High remnant retention fractions will be treated in the discussion of the influence of the retention fraction in Sect.~\ref{sec:remn}.} The range of the dissolution timescale parameter $t_0$ and resulting total disruption times that are considered in Fig.~\ref{fig:mf1} cover two orders of magnitude. As the total lifetime increases, the depletion of the low-mass stellar MF close to total disruption becomes more prominent. Conversely, the MF of short-lived clusters is depleted around $m\sim 1~\msun$. As introduced in the last paragraphs of Sect.~\ref{sec:mf}, this difference is caused by the fixed timescale on which stellar evolution decreases the maximum stellar mass, implying that the masses of the most massive stars are larger in quickly dissolving clusters than in slowly dissolving ones. Because in short-lived clusters the massive stars are still present when the bulk of the dissolution occurs, their rapid two-body relaxation with intermediate-mass stars dominates over the relatively close proximity to the escape energy of low-mass stars, yielding a depletion at intermediate masses. In long-lived clusters, this cannot occur because the very massive stars have disappeared before the mass loss by dissolution becomes important, thus resulting in the depletion of the very low-mass end of the MF. As a rule of thumb, for $t<400$~Myr (which is the lifetime of a $3~\msun$ star) the depletion typically occurs around 15---20\% of the mass of the most massive star (see Sect.~\ref{sec:mf}). In terms of the total disruption time, the transition from intermediate-mass star depletion to low-mass star depletion occurs around $\tdis\sim 1$~Gyr.

\begin{figure}[t]
\resizebox{\hsize}{!}{\includegraphics{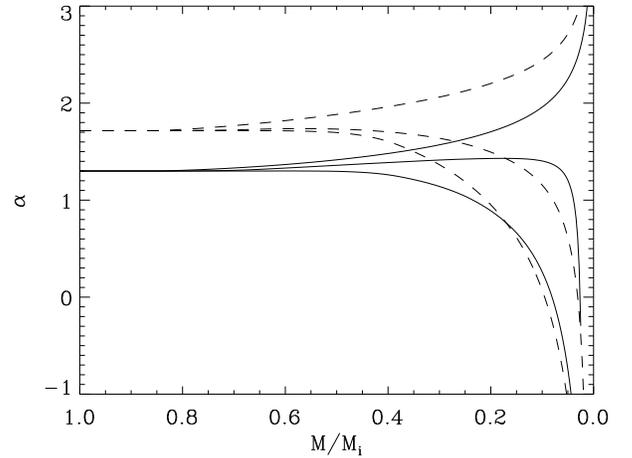}}
\caption[]{\label{fig:alpha1}
       Influence of the disruption time on the stellar MF slope $\alpha$ in the range $0.1<m/\msun<0.5$ (solid) and $0.3<m/\msun<0.8$ (dashed) for a cluster with a low remnant retention fraction ($\log{(M_{\rm i}/\msun)}=4.5$). Shown is $\alpha$ versus the remaining cluster mass fraction. From top to bottom, for each mass range the lines represent $t_0=\{0.1,1,10\}$~Myr, corresponding to $\tdis=\{0.16,1.42,12.26\}$~Gyr.
                 }
\end{figure}
A quantifiable way to look at the evolution of the stellar MF in star clusters is to consider the slope of the MF $n_{\rm s}\propto m^{-\alpha}$ in certain mass intervals \citep{richer91,demarchi07,demarchi07b,vesperini09}. For the commonly used mass intervals $0.1<m/\msun<0.5$ ($\alpha_1$) and $0.3<m/\msun<0.8$ ($\alpha_2$), Fig.~\ref{fig:alpha1} shows the evolution of the slope $\alpha$ for the same clusters as before. Like Fig.~\ref{fig:mf1}, this illustrates that for short disruption times the slope steepens as the cluster dissolves, while for long disruption times the slope flattens with time. The presented models and other model runs indicate that $\alpha_1$ increases with time for $\tdis<1$~Gyr and decreases for $\tdis>2$~Gyr. For total disruption times in between these values, the slope first increases and then decreases. The slope in the second mass interval $\alpha_2$ shows the same behaviour. It increases for $\tdis<0.5$~Gyr and decreases for $\tdis>1$~Gyr.

The mass-to-light ($M/L$) ratio evolution of star clusters is affected by the evolution of the MF due to the large variations in $M/L$ ratio between stars of different masses. Massive stars have lower $M/L$ ratios than low-mass stars, implying that a cluster with a MF that is depleted in low-mass stars will have a reduced $M/L$ ratio \citep{baumgardt03,kruijssen08b,kruijssen08}. As such, one would also expect a correlation between the slope of the MF and $M/L$ ratio. 

\begin{figure}[t]
\resizebox{\hsize}{!}{\includegraphics{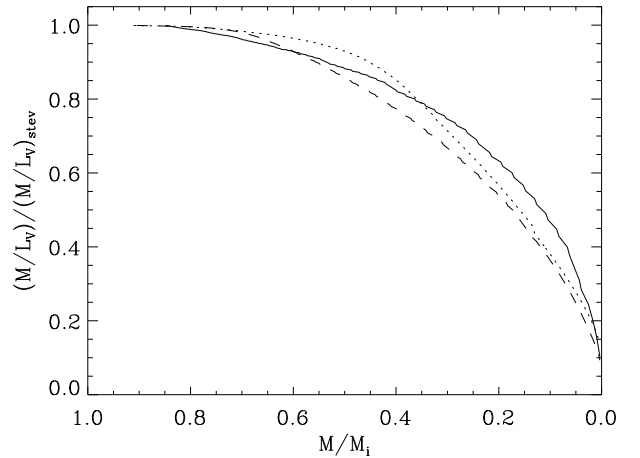}}
\caption[]{\label{fig:ml1}
       Influence of the disruption time on the $M/L_V$ ratio evolution for a cluster with a low remnant retention fraction ($\log{(M_{\rm i}/\msun)}=4.5$). Shown is the relative $M/L_V$ ratio decrease with respect to the value expected for stellar evolution $(M/L_V)_{\rm stev}$ versus the remaining cluster mass fraction. The solid, dashed and dotted lines represent $t_0=\{0.1,1,10\}$~Myr, respectively, corresponding to $\tdis=\{0.16,1.42,12.26\}$~Gyr.
                 }
\end{figure}
In Fig.~\ref{fig:ml1}, the evolution of the ratio of the $V$-band $M/L_V$ to the mass-to-light ratio due to stellar evolution $(M/L_V)_{\rm stev}$ is shown for the same clusters as in Figs.~\ref{fig:mf1} and~\ref{fig:alpha1}. This quantity reflects the relative $M/L_V$ ratio change induced by dynamical evolution with respect to evolutionary fading only. If the ejection rate would be independent of stellar mass, the evolution would follow a horizontal line at $(M/L_V)/(M/L_V)_{\rm stev}=1$. However, when accounting for dynamical evolution, the $M/L$ ratio is always smaller than that for stellar evolution only. Somewhat surprisingly, this is also the case for clusters for which the slope of the MF increases (see Fig.~\ref{fig:alpha1}). This is explained by looking at the evolution of the entire MF in Fig.~\ref{fig:mf1}. Even though the slope at low masses increases for short disruption times due to the ejection of intermediate-mass stars, the most massive stars that dominate the cluster light are still retained. Because stars of intermediate masses are lost instead, the $M/L$ ratio decreases.

\begin{figure}[t]
\resizebox{\hsize}{!}{\includegraphics{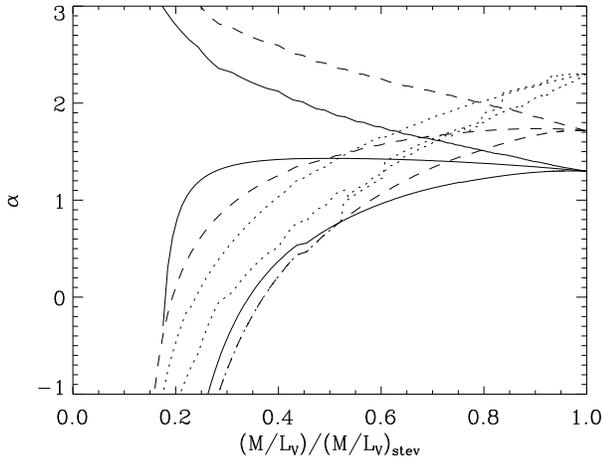}}
\caption[]{\label{fig:mlalpha1}
       Influence of the disruption time on the combined evolution of the MF slope $\alpha$ and the $M/L_V$ ratio for a cluster with a low remnant retention fraction ($\log{(M_{\rm i}/\msun)}=4.5$). Shown is $\alpha$ versus the relative $M/L_V$ ratio decrease due to dynamical evolution. All clusters start at the vertical line $(M/L_V)/(M/L_V)_{\rm stev}=1$. Solid lines denote the slope in the mass range $0.1<m/\msun<0.5$, dashed lines designate the mass range $0.3<m/\msun<0.8$, and dotted lines represent the mass range $0.3m_{\rm max}(t)<m/\msun<0.8m_{\rm max}(t)$, with from top to bottom $t_0=\{0.1,1,10\}$~Myr, corresponding to $\tdis=\{0.16,1.42,12.26\}$~Gyr.
                 }
\end{figure}
Because the slope of the stellar MF either increases or decreases at masses $m<1~\msun$, the decrease of the $M/L$ ratio implies a large range of MF slopes that can occur at low $M/L$ ratios. This is shown in Fig.~\ref{fig:mlalpha1}, where the relation between $\alpha$ and the $M/L$ ratio drop is presented. The slope of the stellar MF in a certain mass range does not necessarily reflect the $M/L$ ratio of the entire cluster. Considering the aforementioned rule of thumb stating that for total disruption times $\tdis<1$~Gyr the depletion of the MF occurs around 15---20\% of the mass of the most massive star $m_{\rm max}(t)$, it is useful to define the slope in a mass range that is related to $m_{\rm max}(t)$. In Fig.~\ref{fig:mlalpha1}, the relation between slope and $M/L$ ratio is also shown for the slope in the stellar mass range $0.3m_{\rm max}(t)<m/\msun<0.8m_{\rm max}(t)$. In such a relative mass range, the slope follows a much narrower relation with $M/L$ ratio. The range between 30\% and 80\% of $m_{\rm max}(t)$ was chosen to maximise this effect.

For the slopes in the fixed stellar mass ranges ($\alpha_1$ and $\alpha_2$, see above), the relation with the $M/L$ ratio becomes better defined for long-lived clusters. It is shown in Figs.~\ref{fig:alpha1}---\ref{fig:mlalpha1} that both the slope and the $M/L$ ratio decrease for clusters with long disruption times, indicating that both quantities are more clearly related for globular cluster-like lifetimes.

\begin{figure}[t]
\resizebox{\hsize}{!}{\includegraphics{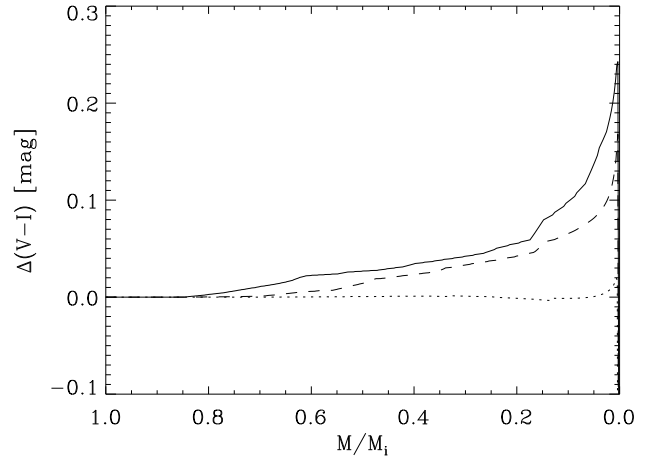}}
\caption[]{\label{fig:colour1}
       Influence of the disruption time on the $V-I$ colour for a cluster with a low remnant retention fraction ($\log{(M_{\rm i}/\msun)}=4.5$). Shown is the colour offset due to dynamical evolution $\Delta(V-I)$ versus the remaining mass fraction. The solid, dashed and dotted lines represent $t_0=\{0.1,1,10\}$~Myr, respectively, corresponding to $\tdis=\{0.16,1.42,12.26\}$~Gyr.
                 }
\end{figure}

The colour of star clusters is also influenced by the evolution of the MF, due to the colour differences between stars of different masses. The $V-I$ magnitude difference $\Delta(V-I)$ with respect to the $V-I$ value that a cluster would have if dynamical evolution were neglected is shown in Fig.~\ref{fig:colour1}. As the clusters dissolve, their colours become redder due to the ejection of main sequence stars. The magnitude difference in $V-I$ exceeds $\Delta(V-I)=0.1$~mag for total disruption times $\leq 1.5$~Gyr. In redder passbands (e.g. the $V-K$ colour), the difference grows to several tenths of magnitudes. For longer total disruption times only stars of the lowest masses are ejected (see Fig.~\ref{fig:mf1}), which hardly contribute to the cluster light and colour, implying that the colours are only marginally affected.

\subsection{The influence of the remnant retention fraction} \label{sec:remn}
The formation of stellar remnants introduces massive bodies in the MF that do not end their lives due to stellar evolution like massive stars do. Depending on their kick velocities, stellar remnants can be retained in (massive) clusters. If they are retained, they keep affecting the evolution of the stellar MF until the cluster is disrupted. Especially black holes can have a pronounced effect on cluster evolution.

\begin{figure}[t]
\resizebox{\hsize}{!}{\includegraphics{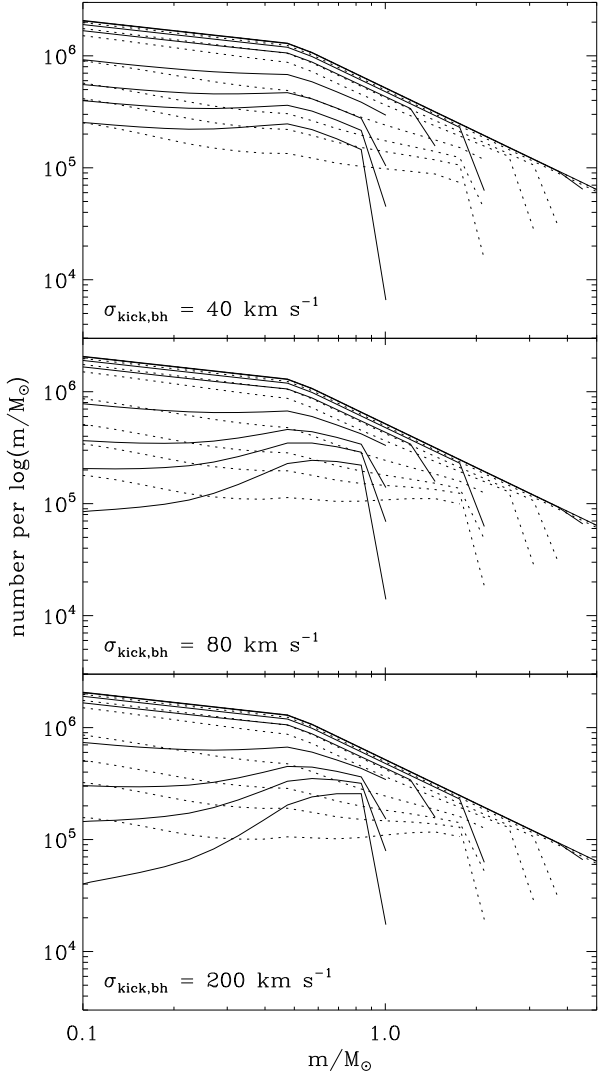}}
\caption[]{\label{fig:mf2}
       Influence of the black hole kick velocity dispersion and disruption time on the evolution of the stellar MF for an initial cluster mass $M_{\rm i}=10^6~\msun$. From top to bottom, the subsequent MFs in each panel are shown for the times at which the remaining cluster mass fraction equals $M/M_{\rm i}=\{1,0.75,0.6,0.5,0.3,0.2,0.15,0.1\}$. Solid lines denote $t_0=1$~Myr ($\tdis=15.13$~Gyr), while dotted lines represent $t_0=0.1$~Myr ($\tdis=1.66$~Gyr).
                 }
\end{figure}

The remnant retention fraction arises from the cluster mass, radius and the kick velocity dispersion (see Eq.~\ref{eq:fret}). {In this section, the mass-radius relation from Eq.~\ref{eq:rh} is used. Although the results will differ for other relations, it has been verified that for commonly used alternatives,\footnote{Such as a constant radius or density.} the change is only marginal and does not affect the nature of the conclusions.} To separate the effect of remnant retention from that of the disruption time, a fixed initial cluster mass of $10^6~\msun$ is assumed while independently varying the velocity dispersion of the remnant kick velocities and the disruption time. The corresponding evolution of the stellar MF is shown in Fig.~\ref{fig:mf2}, for the standard model (see the beginning of this section) with black hole kick velocity dispersions $\sigma_{\rm kick,bh}=\{40,80,200\}$~km~s$^{-1}$, equivalent to $f_{\rm ret,bh}=\{0.219,0.041,0.003\}$ for a $10^6~\msun$ cluster, and for dissolution timescale parameters $t_0=\{0.1,1\}$~Myr, which for a $10^6~\msun$ cluster implies $\tdis=\{1.66,15.13\}$~Gyr. Assuming an age of 12~Gyr, the present-day mass in the case of $t_0=1$~Myr is about $M\sim6\times 10^4~\msun$, comparable to globular clusters. The remaining fraction of the initial mass is $M/M_{\rm i}\sim 0.06$.

If the velocity dispersion of black hole kicks is low and a relatively large fraction of black holes is retained, then the ejection rate of massive stars is increased with respect to high kick velocity dispersions. This arises due to the quick two-body relaxation between the massive stars and the black holes, which will have masses larger than the most massive stars after a few Myr of stellar evolution. As a result, the ejection rate of low-mass stars is largest in clusters containing only few black holes. This happens for clusters with either long or short disruption times, but the effect is largest for long-lived clusters (the solid lines in Fig.~\ref{fig:mf2}). In these clusters the maximum stellar mass is more strongly decreased by stellar evolution than in short-lived clusters, implying that the black hole masses are larger compared to the most massive stars in these clusters. For long disruption times, the presence of massive remnants therefore has a more pronounced effect on the ejection rate of massive stars than for short disruption times. If these long-lived clusters retain a sufficiently large fraction of the stellar remnants, their stellar MF may even become depleted in massive stars.

The top panel of Fig.~\ref{fig:mf2} also shows that for a cluster with a high remnant retention fraction, the impact of the disruption time on the MF evolution is similar to that of clusters with low retention fractions (see Fig.~\ref{fig:mf1}). However, the influence of the disruption time becomes smaller when more remnants are retained. This explains why \citet{baumgardt03} only found a very weak dependence of the evolution of the MF on the disruption time (also see Fig.~\ref{fig:comp}), since they neglected remnant kick velocities and retained all remnants in their simulations.

\begin{figure}[t]
\resizebox{\hsize}{!}{\includegraphics{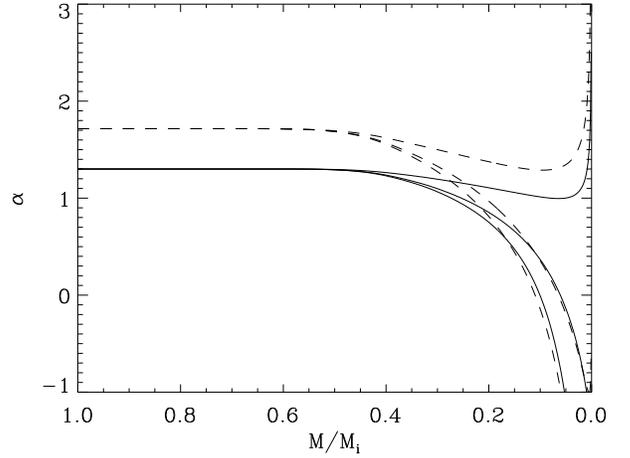}}
\caption[]{\label{fig:alpha2}
       Influence of the black hole retention fraction on the stellar MF slope $\alpha$ in the range $0.1<m/\msun<0.5$ (solid) and $0.3<m/\msun<0.8$ (dashed) for an initial cluster mass $M_{\rm i}=10^6~\msun$. Shown is $\alpha$ versus the remaining cluster mass fraction. From top to bottom, for each mass range the lines represent $\sigma_{\rm kick,bh}=\{40,80,200\}$~km~s$^{-1}$, corresponding to $f_{\rm ret,bh}=\{0.219,0.041,0.003\}$ for a $10^6~\msun$ cluster.
                 }
\end{figure}
Analogous to Fig.~\ref{fig:alpha1} in Sect.~\ref{sec:tdis}, the evolution of the MF slope in different mass ranges is shown in Fig.~\ref{fig:alpha2} for the clusters with $t_0=1$~Myr from Fig.~\ref{fig:mf2}.\footnote{For the clusters with relatively long disruption times that are considered in this section, the variable stellar mass range that was introduced in Sect.~\ref{sec:tdis} to trace the relation between MF slope and $M/L$ ratio gives an evolution of the slope that is comparable that for the fixed mass ranges. It is omitted from the figures in this section to improve their clarity.} The kick velocity dispersion has an effect on the MF that is more uniform than the consequences of variations in the disruption time, leading to very similar slope evolutions in the two different stellar mass ranges. Independent of the mass range, an increase in remnant retention fraction is reflected by an increase of $\alpha$. The model that is displayed for $\sigma_{\rm kick,bh}=40$~km~s$^{-1}$, $t_0=1$~Myr, and $M_{\rm i}=10^6~\msun$ (the upper dashed and solid lines in Fig.~\ref{fig:alpha2}) marks the transition between an increase or decrease of the MF slope by dynamical evolution. For an initial $f_{\rm ret,bh}<0.25$, low-mass stars are preferentially ejected during most of the cluster lifetime, while for $f_{\rm ret,bh}>0.25$ mainly the massive stars escape. For shorter disruption times, the transition is located at a smaller black hole retention fraction.

\begin{figure}[t]
\resizebox{\hsize}{!}{\includegraphics{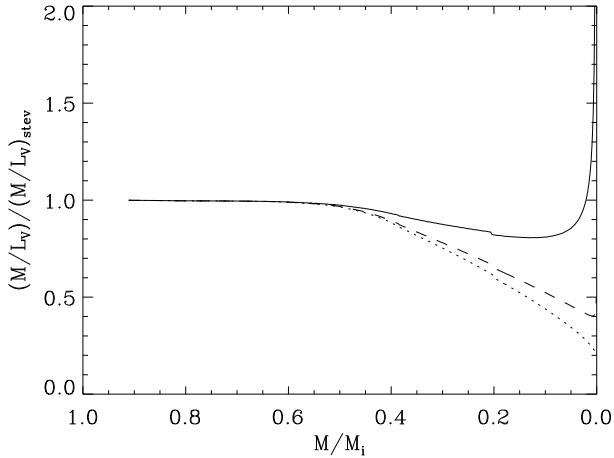}}
\caption[]{\label{fig:ml2}
       Influence of the black hole retention fraction on the $M/L_V$ ratio evolution for an initial cluster mass $M_{\rm i}=10^6~\msun$. Shown is the relative $M/L_V$ ratio decrease with respect to the value expected for stellar evolution $(M/L_V)_{\rm stev}$ versus the remaining cluster mass fraction. The solid, dashed and dotted lines represent $\sigma_{\rm kick,bh}=\{40,80,200\}$~km~s$^{-1}$, corresponding to $f_{\rm ret,bh}=\{0.219,0.041,0.003\}$ for a $10^6~\msun$ cluster.
                 }
\end{figure}

Because the black hole retention fraction affects the overall slope of the stellar MF, the changes in $\alpha$ are matched by corresponding changes in the $M/L$ ratio. In Fig.~\ref{fig:ml2}, the relative $M/L_V$ ratio change due to dynamical evolution is shown for same clusters as in Fig.~\ref{fig:alpha2}. Contrary to the clusters with low remnant retention fractions in Sect.~\ref{sec:tdis}, the $M/L$ ratio of the clusters in Fig.~\ref{fig:ml2} does not monotonously decrease. Close to total disruption, the massive remnants are the last bodies to be ejected. During that short phase of cluster evolution, the $M/L$ ratio is increased by dynamical evolution and exceeds the value it would have due to stellar evolution alone.

\begin{figure}[t]
\resizebox{\hsize}{!}{\includegraphics{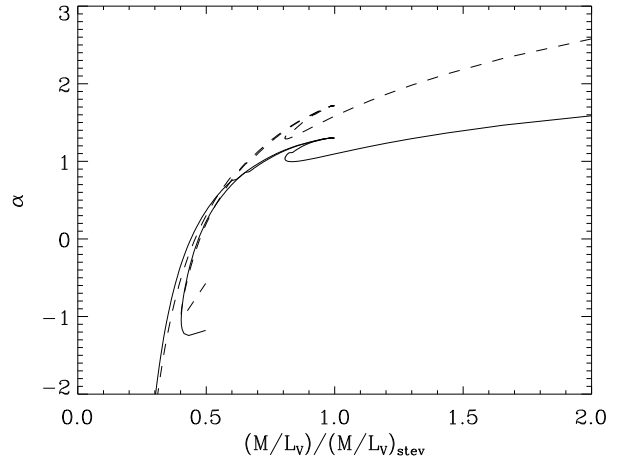}}
\caption[]{\label{fig:mlalpha2}
       Influence of the black hole retention fraction on the combined evolution of the MF slope $\alpha$ and the $M/L_V$ ratio for an initial cluster mass $M_{\rm i}=10^6~\msun$. Shown is $\alpha$ versus the relative $M/L_V$ ratio decrease due to dynamical evolution. All clusters start at the vertical line $(M/L_V)/(M/L_V)_{\rm stev}=1$. Solid lines denote the slope in the mass range $0.1<m/\msun<0.5$ and the dashed lines designate the mass range $0.3<m/\msun<0.8$, with from right to left $\sigma_{\rm kick,bh}=\{40,80,200\}$~km~s$^{-1}$, corresponding to $f_{\rm ret,bh}=\{0.219,0.041,0.003\}$ for a $10^6~\msun$ cluster.
                 }
\end{figure}

The behaviour of $M/L$ ratio for different black hole kick velocity dispersions has interesting implications for the relation between stellar MF slope and $M/L$ ratio, which is shown in Fig.~\ref{fig:mlalpha2}. In combination with Fig.~\ref{fig:mlalpha1} (note the different axes), it shows possible evolutionary tracks of star clusters in this plane, indicating that nearly every location may be reached. However, when limiting ourselves to long-lived clusters, Fig.~\ref{fig:mlalpha2} illustrates that these clusters will follow a trend of decreasing slope with decreasing $M/L$ ratio, albeit with excursions to high $M/L$ ratios and slightly higher $\alpha$ close to their total disruption. This explains the trend that was found by \citet{kruijssen09}, who considered the relation between the observed MF slopes and $M/L$ ratios of Galactic globular clusters.

The colour change due to dynamical evolution is only very small for clusters with $\tdis>1.5$~Gyr (see Sect.~\ref{sec:tdis}). Because clusters in which remnants are retained are massive, their lifetimes are correspondingly long. As a result, the colour evolution is largely unaffected for the clusters in which the remnant retention fraction could play a role ($\Delta(V-I)<0.03$~mag). The colour change is even smaller if more massive remnants are retained, because then the stellar MF more closely resembles its initial form (see the upper panel of Fig.~\ref{fig:mf2}). Long-lived clusters generally appear $\sim 0.005$~mag bluer in $V-I$ due to dynamical evolution during the last $\sim 3$---20\% of their lifetimes and reach a similar reddening upon their total disruption, which is well within observational errors. The colours of old clusters are thus only marginally affected by dynamical evolution.

\begin{figure}[t]
\resizebox{\hsize}{!}{\includegraphics{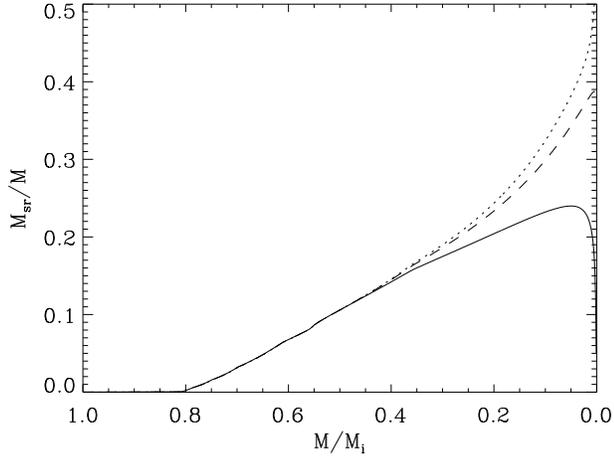}}
\caption[]{\label{fig:mcsr2}
       Influence of the black hole retention fraction on the total remnant mass fraction. Shown is the ratio of the total mass in stellar remnants $M_{\rm sr}$ to the cluster mass $M$ versus the remaining cluster mass fraction. The solid, dashed and dotted lines represent $\sigma_{\rm kick,bh}=\{40,80,200\}$~km~s$^{-1}$, corresponding to $f_{\rm ret,bh}=\{0.219,0.041,0.003\}$ for a $10^6~\msun$ cluster.
                 }
\end{figure}

The evolution of the total remnant mass fraction is shown in Fig.~\ref{fig:mcsr2} for different black hole kick velocity dispersions. The seemingly counterintuitive result is that the fraction of the cluster mass that is constituted by remnants is smaller when more black holes are retained. As shown in Fig.~\ref{fig:mf2}, the retention of black holes suppresses the depletion of the low-mass end of the MF due to the `sweet spot' ejection (see Sect.~\ref{sec:mf}) of massive ($\sim 1~\msun$) stars by the black holes. After $\sim 1$~Gyr, white dwarfs and neutron stars have masses that are similar to those of the massive stars, implying that their ejection rate is also increased when more black holes are retained. Because the total mass constituted by white dwarfs and neutron stars is larger than the combined mass of all black holes, the fraction of the total cluster mass that is constituted by remnants decreases if these low-mass remnants are ejected by the more massive black holes.

\section{Discussion and applications} \label{sec:disc}
The results of this paper show that the stellar MFs in star clusters differ strongly from their initial forms due to dynamical cluster evolution. The specific kinds of these differences depend on the properties of the star clusters and their tidal environment, most importantly on the disruption time, remnant retention fraction, and IMF.\footnote{Although not specifically shown in this paper (but not surprisingly), the differences also depend on the initial-final stellar mass relation.}

\begin{figure}[t]
\resizebox{\hsize}{!}{\includegraphics{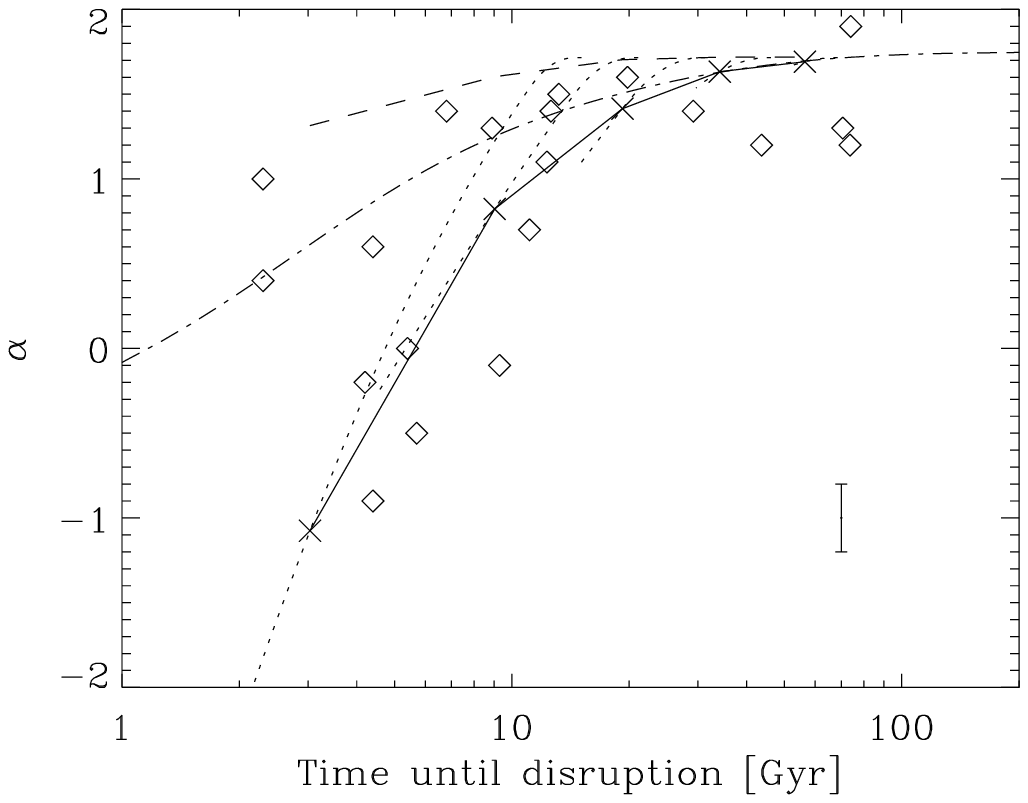}}
\caption[]{\label{fig:alphatdis}
       MF slope versus remaining lifetime (assuming a globular cluster age of 12~Gyr). Diamonds represent the observed values from \citet{demarchi07}, with typical errors as shown by the error bar in the lower right corner. The remaining lifetimes are taken from \citet{baumgardt08b}. Dotted curves represent the model evolutionary tracks of clusters with $\log{(M_{\rm i}/\msun)}=\{6,6.25,6.5,6.75,7\}$ from Sect.~\ref{sec:remn} with $\{\sigma_{\rm kick,wd},\sigma_{\rm kick,ns},\sigma_{\rm kick,bh}\}=\{4,100,200\}$~km~s$^{-1}$, corresponding to $\{f_{\rm ret,wd},f_{\rm ret,ns},f_{\rm ret,bh}\}=\{0.983,0.022,0.003\}$ for a $10^6~\msun$ cluster. The solid line connects the present-day locations of the modeled clusters in the diagram (crosses), while the dashed line represents the same relation for $\sigma_{\rm kick,bh}=40$~km~s$^{-1}$ ($f_{\rm ret,bh}=0.219$ for a $10^6~\msun$ cluster). The dash-dotted line shows the homologous cluster evolution from \citet{baumgardt03}.
                 }
\end{figure}

A physical model for the evolution of the stellar MF is presented in which two-body relaxation leads to a stellar mass dependence of the ejection rate. For any particular stellar mass, the ejection rate is determined by the typical proximity of that mass to the escape energy and by the timescale on which the two-body relaxation with the other stars takes place. Combined with a prescription for stellar evolution, stellar remnant production, and remnant retention using kick velocity dispersions, this provides a description for the total evolution of the MF. {\it This description is independent of the adopted total mass evolution}. {The model shows that the slope of the mass function is a possible indicator for the mass fraction that has been lost due to dissolution, provided that the IMF does not vary and the remnant retention fraction has been fairly similar for young globular clusters.\footnote{Any variability of the retention fraction would induce substantial scatter, see Sect.~\ref{sec:remn} and Fig.~\ref{fig:alphatdis}.}} 

For the exact same initial conditions, the model shows excellent agreement with $N$-body simulations of the evolving MF by \citet{baumgardt03}. However, an important advantage of the presented model compared to the (more accurate) $N$-body simulations is its short runtime and corresponding flexibility. It can be easily applied to compute the evolution of clusters for a large range of initial conditions. The results can then be used to identify interesting cases for more detailed and less simplified calculations with $N$-body or Monte Carlo models.

{The most important simplification of the model is neglecting the effect of binary encounters on the stellar mass dependence of the ejection rate. To incorporate binaries, a conclusive census of the binary population in star clusters would be required, which is not yet available. Nonetheless, it is possible to make a qualitative estimate for the effect of binaries. The encounter rate of binaries would typically be higher than that of individual stars, because the cross section of binaries is larger. This would increase the relative escape rate at the stellar mass for which the binary fraction\footnote{The fraction of stars residing in binary or multiple systems.} peaks. This binary fraction is found to increase with primary mass \citep[see e.g.][]{kouwenhoven09}. Because massive stars are removed by stellar evolution, this implies that the binary fraction decreases with age, which is in agreement with the low binary fraction observed in globular clusters \citep[$\sim 2\%$, e.g.][]{richer04}. The effect of binaries on the evolution of the mass function would thus be most notable if the majority of the dynamical mass loss occurs at ages $< 50$~Myr (the typical lifetime of an $8~\msun$ star), in which case it would somewhat enhance the relative escape rate of the most massive stars.}

The model is applied to investigate the influence of the disruption time and remnant retention on the evolution of the MF and integrated photometric properties of star clusters. For total disruption times $\tdis<1$~Gyr, the modeled relative ejection rate is highest at a certain `sweet spot' mass that is typically 15---20\% of the mass of the most massive objects in the cluster. For longer lifetimes, the evolution of the MF is dominated by low-mass star depletion, unless the retention fraction of massive stellar remnants is larger than 0.25. Only in the particular case of such a high retention fraction, the $M/L$ ratio is increased by dynamical evolution when the cluster approaches total disruption. In all other scenarios, the $M/L$ ratio decreases because the most massive (luminous) stars are kept.\footnote{This process differs from a possible variability of the proportionality between the velocity dispersion and the cluster mass, which concerns a much shorter timescale \citep[e.g.][]{boily09}.} When defining the slope of the MF in the range 30---80\% of the maximum stellar mass, this gives a clear relation between the MF slope and the $M/L$ ratio. For slopes that are defined in fixed mass ranges, there is not necessarily a correlation between slope and $M/L$ ratio if $\tdis<1$~Gyr. In clusters with a longer total disruption time, both quantities are related. Dynamical cluster evolution is found to induce some reddening of the integrated cluster colours, amounting up to 0.1---0.2~mag in $V-I$ for total disruption times $\tdis<1.5$~Gyr. The fraction of the cluster mass that is constituted by remnants surprisingly decreases if more black holes are retained, because the black holes preferentially eject bodies around the masses of white dwarfs and neutron stars, which contain most of the total remnant mass.

Contrary to what is suggested by other studies \citep[e.g.][]{baumgardt03,anders09}, the evolution of the MF is not homologous. The reason that these studies concluded that its evolution is very similar for all clusters (also see Figs.~\ref{fig:comp} and~\ref{fig:comp2}), is that they assumed that all remnants were retained. It is illustrated in Fig.~\ref{fig:mf2} that the differences between clusters with dissimilar disruption times disappear when the retention fraction increases. For realistic retention fractions, differences do arise. If two clusters with different initial masses have the same total disruption time, their MF evolution will be dissimilar due to their different remnant retention fractions and the impact of the retained remnants on the dynamical cluster evolution. Alternatively, if two clusters have equal initial masses but different total disruption times, for instance due to differences in their galactic location or environment, their MF evolution will be dissimilar due to the dynamical impact of the evolution of the maximum stellar mass. 

The larger variation of MF evolution that is found with presented model may also be able to explain observations of globular clusters in which the MF cannot be characterised by a single power law \citep{demarchi00}. If the evolution of the MF were homologous, these features would likely be primordial \citep{baumgardt03}, but this is not necessarily the case when using realistic retention fractions. Most other differences between the results presented in Sect.~\ref{sec:results} and those from \citet{baumgardt03} are also due to their assumption of full remnant retention. For example, their $M/L$ ratio evolution shows a smaller decrease than in Fig.~\ref{fig:ml1}. This is explained in Fig.~\ref{fig:ml2}, where it is shown that dynamical evolution reduces the $M/L$ ratio by a smaller amount if the retention fraction is larger.

{Studies on the fractal nature of cluster formation show that star clusters are initially substructured \citep{elmegreen00,bonnell03}. Even though this substructure is typically erased on a crossing time, it can induce primordial mass segregation in star clusters \citep{mcmillan07,allison09}.} The influence of primordial mass segregation on the evolution of the MF has recently been investigated by \citet{baumgardt08b} and \citet{vesperini09}. While \citet{baumgardt08b} do not include stellar evolution and concentrate on two-body relaxation, \citet{vesperini09} do include stellar evolution. They show that for some degrees of primordial mass segregation, the mass loss by stellar evolution can induce additional dynamical mass loss that strongly decreases the total disruption time. For clusters that survive for a Hubble time, the MF evolution in the case of primordial mass segregation is very similar to an initially unsegregated cluster. \citet{vesperini09} conclude that the evolution of the MF is only affected by primordial mass segregation for clusters in which the total disruption time is sufficiently decreased by the induced mass loss. In that case, the slope of the MF remains much closer to its initial value than it would in clusters without primordial mass segregation. Their conclusion is consistent with the model presented in this paper, because the evolution of the MF is determined by the most massive stars at the time when the largest mass loss occurs (see Figs.~\ref{fig:chinorm} and~\ref{fig:mf1}). This induced mass loss enters the model in terms of the absolute mass loss rate in Eq.~\ref{eq:dmdt}, not in the stellar mass-dependent escape rate per unit mass loss rate of Eq.~\ref{eq:chi}.

A change in total mass loss rate is not the only consequence of primordial mass segregation. \citet{baumgardt08b} have shown that low-mass star depletion is enhanced for clusters without stellar evolution that are primordially mass-segregated. This occurs because energy equipartition is reached on a shorter timescale and because of their use of a fixed ($m_{\rm max}=1.2~\msun$) maximum stellar mass. As a result, there are no massive bodies to increase the ejection rate of intermediate mass stars (see Fig.~\ref{fig:maxchinorm}), implying that only the low-mass stars are preferentially lost. In the present paper, mass segregation is assumed to arise dynamically, but the model could in principle be adapted to cover primordial mass segregation by setting $c_2=0$ and adjusting $c_1$ to the initial velocity distribution until it is erased by dynamical evolution (see Eq.~\ref{eq:tseg}), after which the values from Sect.~\ref{sec:mf} can be used.\footnote{As explained in Sect~\ref{sec:mf}, $c_1$ represents the ratio of the mean speed squared to the central escape velocity squared that depends on the degree of mass segregation (and thus on the IMF). On the other hand, $c_2$ is a proportionality constant in the expression for the onset of the stellar mass-dependent ejection of stars, which depends on the concentration or King parameter.} This does not necessarily yield enhanced low-mass star depletion for clusters with a complete IMF (including masses $m>1.2~\msun$) because of the presence of massive stars or remnants.

The presented model can be applied to the MFs of Galactic globular clusters that are observed by \citet{demarchi07}. These MFs are more strongly depleted than is found in the $N$-body simulations by \citet{baumgardt03}, which has been attributed to primordial mass segregation \citep{baumgardt08b}. However, the observations can also very accurately be explained with the realistic remnant retention fractions that are used in the present paper. This is shown in Fig.~\ref{fig:alphatdis}, where the observed MF slopes and remaining lifetimes of the globular clusters from \citet{demarchi07} are compared with the globular cluster-like models from Sect.~\ref{sec:remn} {($t_0=1$~Myr)}. The models are in much better agreement with the data than the $N$-body runs with complete remnant retention from \citet{baumgardt03}. Deviations to other values of $\alpha$ can occur due to variations in disruption time and remnant retention fractions, as is also shown in Fig.~\ref{fig:alphatdis}. For example, a variation of the remnant kick velocity with metallicity in combination with the known variation of the disruption time \citep[see, e.g.][]{kruijssen09,kruijssen09b} should be sufficient to cover the observed scatter. 

{The above line of reasoning provides an explanation for the the depleted MFs in Fig.~\ref{fig:alphatdis} that is consistent with the simulations by \citet{vesperini09}, who showed that the effects of primordial mass segregation are in fact suppressed in long-lived clusters due to the expansion caused by stellar evolution. This increases the relaxation time and yields an evolution of the MF that is very similar to the initially unsegregated scenario, indicating that primordial mass segregation is not a likely explanation for strongly depleted MFs.} Observations of the remnant composition of these globular clusters could reveal {a definitive answer as to} whether the depleted MFs are explained by primordial mass segregation or by dynamical evolution with a realistic remnant retention fraction.

Dynamical cluster evolution does not appear to have a large effect on the colours of old (globular) clusters. The only way in which the colours could be affected beyond typical observational errors, is if globular clusters have lost substantial fractions of their masses during the first $\sim$~Gyr after their formation. In that case, the dynamical evolution of the stellar MF in globular clusters may have implications for studies of colour bimodality \citep[e.g.][]{larsen01} or the blue tilt \citep[e.g.][]{harris06}. It could then also possibly explain the trend of increasing $V-K$ colour with decreasing $M/L_V$ ratio found by \citet{strader09} for globular clusters in M31, because quickly dissolving clusters generally become redder and have reduced $M/L$ ratios. More research is needed to determine the role of the changing MF in the above properties of globular cluster systems.

It can be concluded that the evolution of the stellar MF in star clusters is not as similar for all clusters as previously thought. Its precise evolution is determined by cluster characteristics like the disruption time, the remnant retention fraction, initial-final stellar mass relation, and the IMF. In order to decipher the evolution of observed star clusters, it is essential to record these characteristics and to relate them to possible scenarios for the internal evolution of clusters. That way, observables like the slope of the MF, the $M/L$ ratio, the broadband colours, and the mass fraction in remnants can be better understood.

\begin{acknowledgements}
I thank Henny Lamers and Simon Portegies Zwart for support, advice and constructive comments on the manuscript. I also thank Douglas Heggie for inspiring suggestions during the KITP conference on globular clusters in Santa Barbara. Holger Baumgardt is acknowledged for interesting discussions and for kindly providing the data from \citet{baumgardt03} and \citet{baumgardt08b}. I would like to thank Nate Bastian, Mark Gieles and Peter Kruijssen for constructive comments on the manuscript. I am indebted to Mark Gieles for kindly providing $N$-body runs for different IMFs. The referee, Thijs Kouwenhoven, is thanked for thoughtful comments. The Institute of Astronomy in Cambridge is gratefully acknowledged for their kind hospitality during my visit, when a large part of this work was carried out. This research is supported by a TopTalent fellowship from the Netherlands Organisation for ScientiÞc Research (NWO), grant number 021.001.038.
\end{acknowledgements}

\bibliographystyle{aa}

\end{document}